\documentclass[12pt,twoside,a4paper]{article}
\newcommand{\dbar}{d\mkern-5mu\mathchar'26}
\def\PARstart#1#2{\begingroup\def\par{\endgraf\endgroup\lineskiplimit=0pt}
    \setbox2=\hbox{\uppercase{#2} }\newdimen\tmpht \tmpht \ht2
    \advance\tmpht by \baselineskip\font\hhuge=cmr10 at \tmpht
    \setbox1=\hbox{{\hhuge #1}}
    \count7=\tmpht \count8=\ht1\divide\count8 by 1000 \divide\count7 by\count8
    \tmpht=.001\tmpht\multiply\tmpht by \count7\font\hhuge=cmr10 at \tmpht
    \setbox1=\hbox{{\hhuge #1}} \noindent \hangindent1.05\wd1
    \hangafter=-2 {\hskip-\hangindent \lower1\ht1\hbox{\raise1.0\ht2\copy1}%
    \kern-0\wd1}\copy2\lineskiplimit=-1000pt}

\usepackage{times}
\usepackage[latin1]{inputenc}
\usepackage[nofoot,margin=25mm,left=25mm,right=25mm,bottom=25mm]{geometry}
\usepackage{multicol}
\usepackage{fancyhdr}
\usepackage{cite}
\usepackage{ccaption}
\usepackage{graphicx}
\usepackage{amsmath}
\usepackage{amssymb}
\newfixedcaption{\figcaption}{figure}

\makeatletter
\renewcommand{\subsection}{\@startsection
  {subsection}
  {2}
  {0mm}
  {4mm}
  {4mm}
  {\bf\normalsize}}
\newcounter{figurecnt}
\newcommand{\mycaption}[1]{\stepcounter{figurecnt}\textbf{Fig \thefigurecnt}:
}
\begin{document}
\thispagestyle{empty}
\noindent
\begin{center}
\begin{Large}
 Ludwig Boltzmann,  Transport Equation and \\
the Second law 
\\[35mm]
\end{Large}
   
\renewcommand{\thefootnote}{\fnsymbol{footnote}} 
 K. P. N.
Murthy \footnote{kpn@igcar.gov.in}
\\[2mm]

 Theoretical Studies Section,\\ 
Materials Science Division,\\
 Indira Gandhi Centre for Atomic Research,\\
 Kalpakkam 603 102,\\ 
Tamilnadu, INDIA\\[35mm]
\end{center}
\begin{abstract}
Ludwig Boltzmann 
had  a hunch that irreversibility exhibited 
by a macroscopic system  arises from  
the reversible dynamics of its microscopic constituents.
He derived a nonlinear  integro-differential equation -
now called the Boltzmann equation -  
for the phase space density of the molecules of a dilute fluid.
He  
showed that the 
Second law of thermodynamics emerges from Newton's
equations of motion. However Boltzmann realized that
stosszahlansatz, employed  in the derivation, 
smuggles in an element of stochasticity into the 
transport equation. 
He then proposed a fully
stochastic description of entropy  which laid the foundation
for statistical mechanics. 
Recent developments, embodied in different  
fluctuation theorems, 
have shown that Boltzmann's hunch
was, in essence,  correct.  
\end{abstract}   
\vskip 2mm
\vskip 5mm
\vfill
\noindent
Based on the 
invited talk given  at the\\
Sixteenth National Symposium 
on Radiation Physics,\\
Meenakshi College for Women, Chennai,\\
 January 18-20, 2006.  \\
See  A. K. Jena, R. Mathiyarasu and V. Gopalakrishnan (Eds.),\\ 
{\it Proc. Sixteenth National Symp. Rad. Phys.},
\\
Indian Society for Radiation Physics
(2006)p1. 
\newpage
\thispagestyle{empty}
\ \ \ \ \ \\[80mm]

\hfill {\bf Everything existing in the Universe is the fruit of chance and necessity}\\[3mm]

\hfill Diogenes Laertius IX
\newpage
\setcounter{page}{1}
\setcounter{footnote}{0}
\thispagestyle{empty} 
\vglue -10mm 
\section{Prologue}
\PARstart{B}{oltzmann}
transport equation has played an important role in 
basic and applied sciences. It is  a nonlinear 
integro-differential equation for the phase space density of the molecules
of a dilute gas. 
It remains today, an important theoretical technique  
for investigating non-equilibrium systems.  
It was derived by Ludwig Eduard Boltzmann (1844 -  1906)
in his  
 {\it further studies on thermal 
equilibrium between gas molecules}  \cite{Boltzmann_1872}, 
published in
the year 1872. Boltzmann 
did this work solely for purpose 
of addressing the conflict between  time-reversal-invariant 
Newtonian mechanics and  time-arrowed 
thermodynamics. Linear version of this  equation 
\cite{linear_transport_equation}
provides an exact description of neutron transport 
in nuclear reactor core and shields. Linear 
transport equation constitutes  
the backbone of nuclear industry. It is indeed appropriate 
that the Indian Society for Radiation Physics (ISRP) has chosen
{\it Boltzmann transport equation} as  focal theme for the 
sixteenth National Symposium on  Radiation Physics  
(NSRP-16), in Meenakshi College for Women, Chennai during 
January 18 - 21, 2006.  The year 2006 marks the hundredth anniversary
of Boltzmann's death.

There are going to be several talks \cite{NSPR_16_LTE}
in this symposium, 
covering various aspects of 
linear transport equation. However, in this opening talk,
 I shall 
deal with  nonlinear transport equation. I shall tell
you of Boltzmann's life-long struggle for comprehending the mysterious
emergence of time asymmetric behaviour of a macroscopic object  from 
the time symmetric behaviour of its microscopic constituents.  
In the synthesis of a  macro from its  micro, 
why and when does time reversal invariance break down? This
is a question that haunted the scientists then,  haunts us now and 
most assuredly shall haunt us in the future, near and far. 

The Second law is about macroscopic phenomena 
being invariably time asymmetric; it is about macroscopic behaviour
being almost always irreversible
\footnote{Deterioration, dissipation, decay and death characterize macroscopic objects
and macroscopic phenomena. A piece of iron rusts; the reverse happens never.
A tomato rots,  inevitably, invariably and irreversibly. An omelet is easily
made from an egg; never an egg from an omelet.

The physicists are puzzled at the Second law. How does it arise ?
An atom - the constituent of a macroscopic object,  obeys  Newton's laws.
Newtonian dynamics  is time reversal invariant. You can not tell the past from
the future; there is the  determinism -  the present holding both,  the 
entire past and the entire future. The atoms, individually obey the 
time reversal invariant Newtonian dynamics; however their collective 
behaviour breaks the time symmetry.  

The philosophers are aghast at the implications of the Second law. 
Does it hold good for the Creator~? 
They are upset at the Second law since it spoils the optimism and 
determinism implicit in for example in the verse below from
Bhagavat Gita, an ancient text from the Hindu Philosophy: 
\begin{center}
Whatever  happened, it happened\\
for good. \\
Whatever is happening, is \\
happening for good.\\
Whatever that will happen, it will \\
be for good. 
\end{center}

Omar Khayyam surrenders to the irreversibility of life when 
he  
writes, 
\begin{center}
The  Moving  Finger  writes;  and, having  writ,\\
Moves  on:  nor  all  your  Piety  nor  Wit\\
Shall  lure  it  back  to  cancel  half  a  Line,\\
Nor  all  your  Tears  wash  out  a  Word  of  it.
\end{center}  
Bernard Shaw, frustrated with the Second law,  exclaims
{\it  youth is wasted on the young}.  Mark Twain hopes fondly for Second law violation when 
he wonders {\it life would be infinitely happier 
if only we could be born at eighty and gradually approach eighteen}.
}.
Physicists think the Second law can not be 
derived from Newton's equations of motion.  
According to them, the Second law 
must be a consequence of our inability to 
keep track of  a large number, typically of the order of $10^{23}$ or more,
 of molecules.
In other words, the origin of the Second law 
is 
statistical. 
It is one thing if 
statistics is used merely  
as  a convenient descriptor of a macroscopic 
phenomenon. It is quite another thing if we 
want to attribute an element of truth to such a description. 
Is it conceivable that nature
is deterministic at micro level and  
stochastic  at macro level? 
Can (microscopic) determinism give rise to (macroscopic) unpredictability?
Boltzmann thought so. 

Boltzmann believed that the Second law is of dynamical 
origin. He proved it through his transport equation and H-theorem. 
At least he thought he did. Several of his fellow  men
thought otherwise. It is this fascinating story of the 
Second law that I am going to narrate to you in this talk. 
I am going  to tell you of the insights that Boltzmann 
provided through his early work on transport equation and 
his later work that laid the foundation for   
Statistical Mechanics - a subject that aims to derive the 
macroscopic properties of matter from the properties of its microscopic 
constituents and their interactions. 
I am also going to tell you of  
nonlinear dynamics and chaos, subjects that have completely 
changed our views about  determinism, dynamics and predictability. 
Now we know  that determinism does not 
necessarily imply predictability. There are a large number of systems that 
exhibit chaotic behavior. Chaos and hence unpredictability is 
a characteristic of dynamics.  Thus, Boltzmann's hunch 
was, in essence, right. It was just that he was 
ahead of his time. 

Boltzmann  staunchly defended the atomistic view. 
He trusted atoms \cite{Cercignani}. He was of the opinion that 
atomistic view helps at least comprehend 
thermal behaviour of dilute fluids. But the most influential 
and vociferous of the German-speaking physics community - the so-called 
energeticists, led by Ernst Mach (1838 - 1916) and
 Wilhelm Ostwald (1853 - 1932) 
did not approve of this. For them, energy was the only fundamental  
physical entity. They dismissed with contempt any attempt to
describe energy or transformation of energy in more 
fundamental atomistic terms or mechanical pictures. 
This lack of recognition from the members of his own 
community allegedly led Boltzmann to commit suicide \footnote{
Boltzmann enjoyed the respect of all his colleagues.
Rejection of his ideas by the energeticists does not seem to
be the only reason  or even one of the reasons  
that drove him to his tragic end. 
Men like myths. Men like heroes. Scientists are no exception.
Scientists need  
heroes - tragic or otherwise. Boltzmann is one such. 
}.
Ironically, Boltzmann died at the dawn of the victory of the atomistic view. 
For, in the year 1905, Albert Einstein (1879  - 1955) 
established unambiguously
the reality of atoms and molecules in his work \cite{Einstein}
on Brownian motion. 
 
\section{{\bf On the nature of things}}
It all started with our efforts to understand the nature of matter, in general and 
of heat, in particular. Ancient man must have definitely speculated
on the possibility of tiny, invisible and indivisible 
particles assembling in very large numbers 
into a  visible continuum of  solids and liquids
 and  an invisible continuum of 
air that surround us. The Greeks had a 
name for the tiny particle: {\it atom} - the uncuttable.
According to Leucippus (440 B.C.) and his student Democritus (370 B.C.) 
atom moves in void, unceasingly and changing course upon collision with another
atom. 
Titus Lucretius Carus (99 B.C. - 55 B.C.) 
mused on the nature of things~
\footnote
         {Lucretius wrote a six books long poem called {\it De Rerum Natura} 
         (On the Nature of Things) on atomism. He writes of
           \begin{center}
            \vglue -6mm
            clothes hung above a surf-swept shore\\
               grow damp; spread in the sun they dry again.\\
          Yet it is not apparent to us how\\
         the moisture clings to the cloth, or flees the heat.\\
          Water, then, is dispersed in particles,\\
                 atoms too small to be observable..... 
                \end{center}
                            }.
According to him all the phenomena we see around are caused by 
invisible atoms moving hither and thither~
\footnote{ 
 The atoms are 
\begin{center}
\vglue -5mm
... shuffled and jumbled
in many ways, in the course\\
 of endless time they are buffeted, driven along\\
 chancing upon all motions, combinations.\\
At last they fall into such an arrangement\\
 as would create this universe....
\end{center}
}.
There was no role for God in his
scheme of things. Atomism of the very early times was  
inherently and fiercely atheistic.
Perhaps this explains why it lost favour and 
languished into oblivion for several
centuries. 
\section{Revival of Atomistic view}
The revival came with the arrival of Galileo Galilei (1564-1642) who wrote
in the year 1638, of the air surrounding the earth and 
of its ability to stand thirty four 
feet of water in a 
vertical tube closed at the top with the open bottom end 
immersed in a vessel of 
water. He also knew of air expanding upon heating and
invented a water-thermo-graph (thermometer). A few years later, his student 
Evangelista Torricelli (1608-1647) correctly concluded
of air pressure and surmised that mercury, fourteen times heavier, 
would rise in the tube only upto thirty  inches. He showed it 
experimentally. Blaise Pascal (1623 -1662) was quick to point out that
Torricelli's reasoning would imply that the pressure of air 
on top of a mountain should be less, which was also verified 
through experiments in 1648. 
Daniel Gabriel Fahrenheit (1686 - 1736) invented the 
mercury thermometer  and 
the Fahrenheit scale of temperature in the year 1714. Andres Celsius (1701-1744) invented the 
centigrade or Celsius scale of temperature in the year 1742.  
Robert Boyle (1627 - 1691) carried out numerous experiments
on the static and kinetic nature of air pressure and showed that 
the product of pressure and volume of a given amount of air remains constant
if the temperature is also kept constant. This is called Boyle's law
\footnote{Boyle got the idea from the paper of Richard  Townley (1638 - 1707)
describing the work Townley carried out with Henry Power, see S. G. Brush
\cite{Brush_1976} Book 1; p.12.}.
Boyle modeled air as a collection of springs that resist compression (which explains
air - pressure) and expands into available space \cite{Boyle}.
Guillaume Amontons (1663 - 1705) experimented on expansion of 
gases with increase of temperature under constant pressure. He 
proposed an absolute zero of temperature  at which,  
volume of a gas becomes zero at constant pressure or the pressure becomes zero
under constant volume. The absolute zero temperature calculated from 
Amontons' experiments turned out to be $- 248\ {}^{\circ}$~C.
But nobody took notice of the Amontons' suggestion of an
absolute temperature scale and absolute zero of temperature
\footnote{A century later, 
Jacques Alexandre C\'esar Charles (1746 - 1823)
and Joseph Louis Gay-Lussac (1778 - 1850) established the 
law of thermal expansion of gases as we know of it today: 
The pressure (at constant volume) or the volume (at constant
pressure) is proportional to $T+\alpha$, where $T$ is 
the temperature measured in some scale say Fahrenheit 
or Celsius; $\alpha$ is a constant 
that  depends on the scale chosen for $T$. 
We can define $T+\alpha$ as absolute temperature whose zero
will be lowest attainable temperature;  in the scale chosen for measuring
$T$ the lowest attainable temperature is thus $-\alpha$.
In fact the notion of absolute scale 
and absolute zero of temperature 
got the acceptance of the scientific community only after 
William Thomson (Kelvin) proposed it \cite{Kelvin}
in the year 1848 based on Carnot engine {\it i.e.} the Second law: 
the partial derivative of 
entropy with respect to energy gives the inverse of absolute temperature.}.  
Another  
important work carried out in the early  eighteenth century
was that of Daniel Bernoulli (1700 - 1782), 
who gave a derivation of Boyle's law from his billiard ball atomic model
\cite{Bernoulli}. Bernoulli's billiard ball atom  moves freely in space,
 colliding with 
other billiard ball atoms and with the walls of the container.
Bernoulli interpreted gas pressure as arising due to numerous impacts 
the billiard ball atoms make with the walls of the container. 
  
\section{Caloric Theory}
Despite these remarkably insightful work, both experimental 
 and theoretical, carried out in the seventeenth and early eighteenth
century, kinetic theory did not take off. 
Scientists could not simply comprehend heat as arising
out of atomic motion: be it undulating motion around fixed position,
like Boyle imagined
 or free motion 
in the available space of the container, like Bernoulli modeled. 
This difficulty is perfectly understandable since it was known
that heat could be transmitted through vacuum,  
like for example,
the heat from the sun. Hence, heat can not be a property of a substance;
it has to be a substance by itself. 
Antoine Lavoisier
(1743 - 1794) gave the name Calorique (or in English Caloric) 
to this fluid substance.
In fact the French chemists included Calorique as one of the elements in the
list prepared in the late eighteenth century.  
Caloric fluid always 
flowed from higher to lower temperatures.  
Heat engines that produced locomotion  
from burning of coal started dotting the European country side.  

\section{Carnot's Engine and Caloric heat}
Nicolas Leonard Sadi Carnot (1796 - 1832) 
was
investigating why a French heat engine 
delivered invariably less work than its British counterpart. Carnot 
was intrigued by the  very idea of  a  heat engine  which manages to do what 
even the almighty Nature could not: 
A heat engine  converts heat into movement. In nature you find that
it is the movement which due to friction 
generates heat and  never the   
the other way. There is no phenomenon like {\it un-friction} 
or {\it anti-friction} which would spontaneously re-assemble the 
heat back into a movement. 
Thinking along these lines Carnot came to the conclusion  \cite{Carnot}
that mere {\it  production of heat is not sufficient to give birth to the 
impelling power; it is necessary there should be cold; 
without it,  heat is useless}. 
Thus the work produced should depend 
on the temperature difference between
the boiler (heat source) and the radiator (the heat sink). This is
a remarkable finding.   
The heat engine is like a mill wheel. A mill wheel simply extracts  work from 
falling water. Larger the quantity of water and higher the fall, 
 more is the 
work produced in the mill wheel. Analogously, larger  the heat
source and higher the temperature fall,   
more is the work 
produced in the heat engine.
If a certain quantity $q$ of caloric falls from absolute temperature $T_1$
to zero, then the work produced will be $W=q$; since it falls only 
to a finite temperature $T_2$ ($0 <T_2 < T_1$),  only the  proportional 
fraction of $q$ 
should equal the work produced. 
In the year 1824, Carnot announced in his historic 
treatise \cite{Carnot}
entitled,  {\it Reflexions on
the motive power of fire and on machines to develop that 
power},  that the ratio of work (W) delivered by a heat engine 
to the heat (q) 
generated in the boiler at temperature $T_1$,   
is given by
\begin{eqnarray}\label{Carnot_efficiency}
\eta = \frac{W}{q} =  
\frac{T_1 - T_2}{T_1}\ < 1\ \  {\rm for}\ \ 0 < T_2 < T_1 < \infty\ ,
\end{eqnarray}
where 
$T_2$ is the temperature of the heat  sink (the radiator).
Even ideally, a heat engine can not have unit efficiency. 
The best
you can get  is  Carnot's efficiency
given by
Eq. (\ref{Carnot_efficiency}). 
 When Carnot measured the 
actual work delivered by a heat engine it was much less than what 
his formula suggested. Real-life heat engines have moving parts 
that rubbed against each other and against  other parts; 
the resulting friction 
- which  produces heat from work - is thus completely
antagonistic to the heat engine which is trying to produce work from 
heat. Not surprisingly a practical engine is less efficient than 
Carnot's ideal engine. In fact Carnot's engine is a double
idealization: its efficiency is less that unity since it is not 
 realistic to set $T_2$ to zero; it should also work without 
friction which is not practical either.    

Carnot's picture of a heat engine is completely consistent 
with the Caloric theory of heat. In fact it constitutes a triumph of the 
Caloric theory.  Water that rotates the mill wheel
is never consumed. Likewise the Caloric fluid that powers the 
 heat engine is never
destroyed. In the radiator the Caloric fluid is reabsorbed in the 
water and returned to the boiler for conversion to steam again. 

It looked like the Caloric theory had come to stay for good. 
It was becoming  immensely  and increasingly 
 difficult for the kinetic heat to dethrone
the Caloric heat and regain its lost and forgotten glory. 
A  sense of complacency  
started prevailing amongst the 
scientists at that time.  
There arose  a certain reluctance to accept new
ideas. It often happens in science: when a scientific theory 
is confirmed and firmly established, it loses its character and becomes a 
dogma; the practitioners of the theory become dogmatic. 

\section{The tragedy of Herpath and Waterston}
Consider the manuscript of 
John Herpath (1790-1868) submitted in the year 1820,
 containing new ideas on kinetic theory
of heat. Herpath, unaware of Bernoulli's work, proposed an atomic
model for the gas; he said heat is proportional to total momentum of
the molecules of gas and absolute temperature corresponds to momentum per 
gas molecule. Herpath's work  was found to be too speculative. 
The Royal Society did not find it
fit to publish it in their Philosophical Transactions. Obviously the 
reviewers were also unaware of the work of Bernoulli.  

The same fate awaited the 
brilliant work scripted by John James Waterston (1811 - 1883), then
 at Bombay (now called Mumbai) and submitted in the year 1845,
 to the Royal Society. Waterston's  model of gas
contained molecules moving incessantly and  colliding 
with each other and with the walls
of the container. Waterston correctly identified the temperature as 
measuring the  energy of motion of the molecules. One of the two reviewers 
considered Waterston's work as \lq~nothing but nonsense~\rq~.
The other reviewer was less harsh. He wrote that  Waterston's
 suggestion that the pressure is due to  
molecular impacts on the walls of the container was \lq~extremely hypothetical
 and difficult to admit~\rq~. The manuscript was rejected 
and buried in the archives of
the Royal Society. 
        Much later, in the year 1891, 
John William Strutt (Lord)  Rayleigh (1842 - 1919)
 stumbled on the 1845-manuscript of Waterston; 
to his astonishment he found it contained essentially the same ideas proposed 
by August Karl   Kr\"onig (1822 - 1879) 
in 1856 \cite{Kronig} and  
by Rudolf Julius Emmanuel Clausius (1822 - 1888) 
in the year 1857 and in the later years. He got Waterston's manuscript
published \cite{Waterston} in the Philosophical 
Transactions of the Royal Society, 
in the year 1893. 

\section{Experimental evidence against Caloric heat}
But then there were significant developments
in experimental thermal physics that 
started knocking at the  very foundations of the Caloric theory.

In the year 1798,  Benjamin Thompson Rumford  (1753 - 1814) 
noticed \cite{rumford} that a canon became hot 
while boring. The heat it generated was sufficient to melt the canon. 
This means that the Caloric fluid produced is 
more than what was 
originally contained in the canon. 
This is not possible under  Caloric theory. 
Julius Robert von Mayer (1814 - 1878), 
in the year 1840,  came to the same conclusion \cite{mayer} that heat is 
like mechanical energy.
The paddle wheel experiment of  James Prescott Joule  (1818 - 1889) 
\cite{joule} carried 
out in the year 1845 established the mechanical equivalence of heat
\footnote{1 Calorie  = 4.184 Joules where Joule is the SI unit of energy 
denoted by the symbol J and given by, 1 J=1Kg. M$^2$/sec$^2$.}. 
These experiments of Rumford, Mayer and Joule, thus  
established unambiguously that the Caloric theory
of heat was wrong and heat, like work,  is actually energy or more precisely
energy in transit, see section \ref{microscopic_heat_and_work}.
Once we identify heat with energy, Carnot's finding becomes intriguing.
Why?

The first law of thermodynamics 
\footnote{This principle of conservation of energy, called
the first law of thermodynamics,  was proposed 
independently by several scientists  
in the middle of the  nineteenth century,
 notable amongst them are Mayer \cite{mayer},
Joule \cite{joule,Joule_first_law} and Helmholtz \cite{Helmholtz}.}
 tells us, energy can 
neither be created nor destroyed.  
However energy  can be converted 
from one form to the other. Carnot's finding amounts to saying that
 heat energy can not
be converted completely into mechanical energy whereas mechanical energy
can be completely converted into heat \footnote{Ginsberg's restatement of the 
three laws of thermodynamics:
\begin{eqnarray}
{\rm  First\  law} &:& {\rm  You\  can't\   win;}\nonumber\\ 
{\rm Second\  law} &:& {\rm You\  can't\  even\  break\  even};\nonumber\\
{\rm  Third\  law} &:& {\rm  You \ can't \  even\  quit.}\nonumber
\end{eqnarray}
}
There is a kind of
 thermodynamic irreversibility.  In the (first-) law abiding
democratic society of energies, 
heat occupies a special place. Perhaps it is like what Bernard Shaw said:
{\it In a democracy, all men are equal but 
some are more equal than others.}
There is an apparent  injustice in nature's scheme. 

Nobody took notice of Carnot's
work for over two decades.
Benoit Paul Emilie Clapeyron (1799 - 1864)
felt that Carnot had discovered something profound.  
He provided the required physical and  mathematical
scaffolding \footnote{The isotherms and
the adiabats in the pressure - volume phase
diagram (describing Carnot's engine)
that you find in text books on thermodynamics 
were actually drawn by Clapeyron} \cite{Clapeyron_1834}
which caught the attention of  
William Thomson (Kelvin) (1824 - 1907) and
Clausius. Kelvin proposed \cite{Kelvin} an absolute 
temperature scale based on Carnot's 
engine.  

\section{Clausius invents  Entropy}
Clausius was intrigued by Carnot's finding. He felt that Carnot's
basic conclusion is  correct and also considered it as of 
great fundamental importance. He called it the Second law of thermodynamics.
But then he rejected Carnot's reasoning  based on Caloric theory of heat. 
From the experiments of Rumford, Mayer and Joule, he understood that
heat and work are simply two different forms of energy transfer, see 
section \ref{microscopic_heat_and_work}. 
He had known by then that heat was a kind 
of motion \cite{Clausius_1857}.
To explain  Carnot's finding in the context of  this 
emerging picture,
Clausius invented a new thermodynamic variable.
His reasoning was simple. 

Consider a thermodynamic process 
described by a path in an appropriate phase space of thermodynamic
variables like internal energy ($E$), volume ($V$), pressure ($P$),
temperature ($T$),
number of molecules ($N$), chemical potential ($\mu$) {\it etc}.
During the process, the system absorbs or liberates energy in the 
form of heat ($Q$) and/or work ($W$). Both $Q$ and $W$ are  path-dependent. 
Hence they  are not  state variables. 
In other words $\dbar Q$
and $\dbar W$ are not  perfect differentials \footnote{The cross
on $d$ 
denotes they are not perfect differentials.}. 
However 
$\dbar W = P dV$. Inverse of pressure provides  integrating
factor for work. Clausius discovered that inverse of temperature
provides   integrating factor for heat. The quantity $\dbar Q/T$ turned 
out to be a perfect differential. Clausius denoted this 
perfect differential by the symbol $dS$.  There was no known
thermodynamic state variable,  
whose perfect differential corresponded 
$dS$. Clausius, in his 
1865 paper \cite{Clausius_1865},
named the state variable $S$ in $dS=\dbar q/T$ as  {\bf entropy} 
\footnote{in the words of Clausius {\it  ... We now seek 
an appropriate name for $S$. .... We would call $S$ 
the transformation content of the body. 
However I have felt it more suitable to take names of 
important scientific quantities from the ancient 
languages in order that they may appear 
unchanged in all contemporary languages. 
Hence I propose that we call $S$
the {\it entropy} of the body after Greek word 
\ \lq\  $\eta\tau\rho o\pi\eta$, meaning 
\lq\lq\ transformation\ \rq\rq\ . I have intentionally 
formed word {\it entropy} to be as similar as possible 
to the word {\it energy}, since
the two quantities that are given these names 
are so closely related
in their physical significance that a 
certain likeness in their names 
has seemed  appropriate.}}.
Let me quickly illustrate this on a simple example.

Start with the first law of thermodynamics,
\begin{eqnarray}
dU = dQ + \dbar W
\end{eqnarray}
Consider an ideal gas to which energy in the form heat is  supplied
at constant volume; its internal  energy increases by $dU=C_V dT$, where 
$C_V$ is the specific heat at constant volume. 
The ideal gas law is given by $PV=\Theta T$, where $\Theta$ is a constant.
From this we get 
\begin{eqnarray}
dV=\Theta\left[ \frac{1}{P}\ dT-\frac{T}{P^{2}}\ dP\right] =0.
\end{eqnarray}
The work done ($-PdV =0 $) is
given by
\begin{eqnarray}\label{work_done}
\dbar W = \Theta\left[ \frac{T}{P}\  dP - dT\right] = 0\ .
\end{eqnarray}
Therefore we have,
\begin{eqnarray}\label{expression_dQ}
dQ=(C_V +\Theta)\  dT - \Theta\  \frac{T}{P}\  dP\ .
\end{eqnarray}
Let us investigate if $dQ$ is a perfect differential. From the above,
we have
\begin{eqnarray}
\frac{\partial Q}{\partial T} & = &C_V +\Theta\ ,  \\
        & &            \nonumber\\
\frac{\partial Q}{\partial P} & = & -\Theta\ \frac{T}{P}\ .
\end{eqnarray}
Differentiating once more  we get,
\begin{eqnarray}
\frac{\partial ^2 Q}{\partial P\partial T}  &=&  0\\ 
& & \\
\frac{\partial ^2 Q}{\partial T\partial P}\ &=&-\frac{\Theta}{P}.
\end{eqnarray}
Therefore,
\begin{eqnarray}
\frac{\partial ^2 Q}{\partial P\partial T}  \ne 
\frac{\partial ^2 Q}{\partial T\partial P}\ ,
\end{eqnarray}
showing that 
$dQ$ is not a perfect differential and 
$Q$ is not a state function of $P$ and $T$.
We shall 
cross the \lq$d\thinspace$\rq\  to denote this.

Consider now,  the quantity $dS=\dbar Q/T$, obtained by dividing
all the terms in Eq. (\ref{expression_dQ}) by $T$.
We have,
\begin{eqnarray}
\frac{\partial S}{\partial T} & = &\frac{C_V +\Theta}{T}\ ,  \\
        & &            \nonumber\\
\frac{\partial S}{\partial P} & = & -\frac{\Theta}{P}\ .
\end{eqnarray}
 It is 
easily seen  that 
\begin{eqnarray}
\frac{\partial ^2 S}{\partial P\partial T}= 
\frac{\partial ^2 S}{\partial T\partial P}=0\ ,
\end{eqnarray}
demonstrating that $S$ is a state function. Clausius
gave the name entropy to this state
function
. 
Thus for the very definition of entropy, we need a 
thermodynamic process that can be represented by a path 
in the space of state variables. We call this a 
quasi-static   process, described below.
\subsection{Quasi-static processes}
Consider a macroscopic system in equilibrium {\it e.g.} a gas in a 
cylinder fitted with a piston. Let $\Lambda$ be a degree of freedom 
which can be manipulated from
outside. For example $\Lambda$ can be taken as volume of the 
gas which can be changed by moving the piston. Consider a 
thermodynamic process in which we switch the value of 
$\Lambda$ from say $\Lambda_0$ to $\Lambda_{\tau}$ 
over a duration of time $\tau$. This switching can 
 be carried out with some  pre-determined protocol. 
For example we can  
change $\Lambda$ uniformly.  
 We say the process 
becomes quasi-static when  the switching takes place
extremely slowly. Strictly for a quasi-static process 
$\tau$ equals infinity. It is a process of slow 
stepping through equilibrium states.
At each step the state variables assume the 
relationship given by equation of states; the system is sort of
dragged through a set of dense succession of equilibrium states. A 
quasi-static process can not be realized in practice. At best
we can approximate it by an extremely slow process. A quasi-static process
is reversible if it takes place at constant total entropy. In other words
during a quasi-static process the change in entropy of the system
plus the change in the entropy of the surroundings equals zero. 
\subsection{The Second law: {\boldmath $dS \ge 0$}}
For defining entropy, Clausius considers a quasi-static reversible process. 
During the process the system absorbs a quantity $\dbar Q_{{\rm rev}}$ of 
reversible heat, from a heat source
at temperature $T$. The entropy of the system increases by
an amount given by,
\begin{eqnarray}\label{Claussius_entropy}
dS=\frac{\dbar Q_{{\rm rev}}}{T}.
\end{eqnarray}
Since the process is quasi-static and reversible, the entropy of the 
heat source decreases by precisely the same amount so that 
the total change in entropy
is zero
\footnote{For the definition of entropy the reversibility 
of the quasi-static process
is only a sufficient condition but not necessary; 
the necessary condition is that the process should be quasi-static.}. 
 
Consider an isolated system and let $dS$ denote the change in entropy
of the system during a process. 
If the system is not isolated, then $ds$  denotes the change in entropy
of the system plus the change in entropy of its surroundings.  
Clausius states the Second law as, 
\begin{eqnarray}
dS \ge 0,
\end{eqnarray}
in any thermodynamic process. 
In the above, equality obtains when the process is quasi-static
and reversible. 
With this Second law assertion,  Clausius was able to show that 
the efficiency of any heat engine is less than or equal to 
that of Carnot's engine, see below.
\section{Kinetic heat replaces Caloric heat  in Carnot's engine}
Consider an engine, M,   which, operating in a cycle, 
draws a quantity $q_1$ of energy in the form of heat  
quasi-statically and reversibly  from a heat reservoir (R) 
at temperature $T_1$. 
Let us say  the engine  converts the entire heat $q_1$ into work
$W$ and returns to its initial state. Such an engine is called
a perfect engine, shown in Fig. (1). 
Under 
the first 
law of thermodynamics
it is 
possible, in 
principle, 
to 
construct 
a perfect engine. 
Let us investigate what happens when we impose the 
Second law. 
The change in entropy of the 
heat source is $-q_1/T_1$. 
Since the engine  returns to its initial
thermodynamic state there is no change 
in its 
entropy. 
We just saw that entropy is a state function.
Thus,  the total change in the entropy 
is 
$dS = -q_1/T_1$. 
The Second law demands that $dS \ge 0$. 
The machine can not deliver work.  Second law forbids perfect engines.
However, the engine can convert 
mechanical energy $W$ 
completely 
into
heat, since during such a  process $dS > 0$.

Consider now an ideal engine M, 
shown in Fig. (2). It draws a quantity 
$q_1$ of energy in the 
form of 
heat, 
quasi-statically and   reversibly 
from a  source, R,  kept at temperature $T_1$; it
converts a part of it into work;  it  
junks the remaining
 part $q_2 < q_1$ 
into a  sink (S), kept at temperature $T_2 < T_1$; then it  returns to 
the state it started with.  
\begin{center}
\begin{figure*}[thp]
\hglue 50mm
\includegraphics[height=64mm,width=65mm]{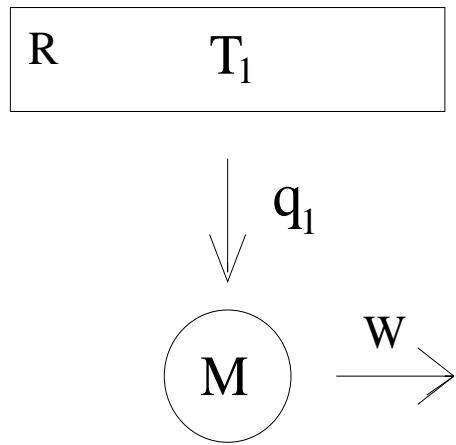}
\caption{Perfect Engine} 
\end{figure*}
\end{center}
\vglue -10mm
\begin{center}
\begin{figure*}[htp]
\hglue 50mm
\includegraphics[height=88mm,width=60mm]{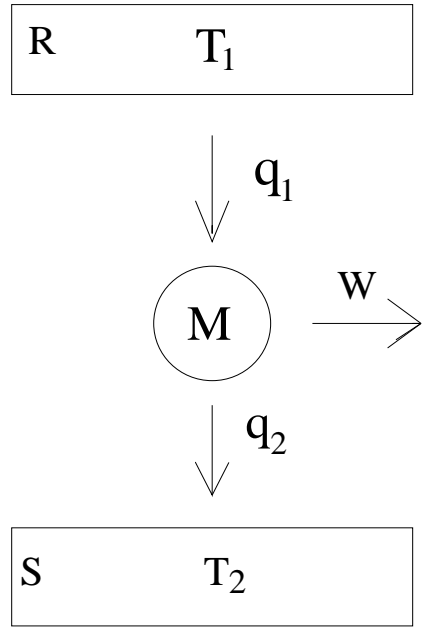}
\caption{Ideal Engine} 
\end{figure*}
\end{center}
\vglue -08mm
\noindent
\noindent

From the first law
we have $q_1-q_2=W$. The efficiency of the engine is 
given by,
$\eta = W/q_1 = 1-(q_2/q_1)$. 
The change in entropy of the heat 
source is 
$-q_1/T_1$ and  that of the  sink is  $q_2/T_2$.
Since the machine returns to its initial state its entropy
does not change. 
Therefore we have, 
\begin{eqnarray}
dS=\frac{q_2}{T_2}-\frac{q_1}{T_1}.
\end{eqnarray}
The Second law demands that $dS\ge 0$. For an ideal 
engine $dS=0$.
Therefore, for an ideal engine  $q_2/q_1=T_2/T_1$,
from  which we get $\eta = 1-(T_2/T_1)$, a result identical to what
Carnot obtained for his Caloric fluid,  
see Eq. (\ref{Carnot_efficiency}). 

This is precisely how the Caloric heat became kinetic heat in
Carnot's engine. If heat is kinetic {\it i.e.} 
motion, then what kind of motion is it?
It must be the irregular motions of the gas molecules;
for, the regular motion is already described by work given in terms of
pressure and change of volume. Entropy, which is heat divided by temperature
must be a measure of this irregularity of molecular motions; a measure of
disorder; a measure of randomness.  
James Clerk Maxwell (1831 - 1879) 
 asserted that the very Second law that talks of increasing 
entropy, must be statistical in character;  hence
it should be possible to contravene the Second law 
with  non zero probability. 
He 
even proposed a
demon - now called Maxwell's demon - that violates the Second
law~\footnote{For an interesting account of Maxwell's demon  
and other demons, see \cite{demon}.}.
For  Maxwell,  
stochasticity  was intrinsic to macroscopic behaviour~
\footnote{
Maxwell was amongst the first to recognize
 the need for statistical approach to kinetic theory. In fact his derivation of the 
distribution of speed of the molecules of an ideal gas 
is ingenious and elegant, see \cite{Maxwell}. He assumes that the 
three components $(v_1, v_2, v_3)$, of the velocity $\vec{v}$ of an 
ideal gas molecule are independent and identically distributed: 
$f(v_1,v_2,v_3)=f(v_1)f(v_2)f(v_3)$, where $f$ is the 
density of molecules in the velocity space. He  argues that since 
there is no preferred direction of motion the function $f$ 
must depend only on  
$v_1^2 + v_2^2 +v_3^2$; this leads to functional equation:
$f(v_1)f(v_2)f(v_3)=\phi(v_1^2+v_2^2+v_3^2)$ whose solution is the 
famous Maxwell-Boltzmann distribution of molecular speed,
\begin{eqnarray}
f(v)=4\pi\bigg(\frac{2\pi k_BT}{m}\bigg)^{-3/2}v^2\exp
\bigg[-\frac{mv^2}{2k_BT}\bigg]\nonumber
\end{eqnarray}
that we are all familiar with.
In the same paper \cite{Maxwell}, Maxwell correctly recognizes that the 
Maxwell-Boltzmann distribution
is a simple consequence of Central Limit Theorem  
concerned with additive random variables,
see footnote~(\ref{clt}).}.
\section{Boltzmann Transport Equation}
However Boltzmann, at least in his early years,
felt there was no need to invoke statistics 
to 
comprehend the Second law. At the age of twenty two, Boltzmann
wrote {\it on the mechanical meaning of the Second law of the theory of heat}
\cite{Boltzmann_1866}.
Of interest to us
is Boltzmann's paper \cite{Boltzmann_1872}
 published in the year 1872 
in which he derived  
his  transport equation and in which he also announced 
the  $H$ theorem to prove 
the Second law.
Boltzmann considers the density  $f(\vec{r}, \vec{p}, t)$ 
of molecules, each of mass $m$, 
at the six-dimensional phase space
point~\footnote{Classically a particle is specified by three position and three 
momentum coordinates. It is represented by a point in the 
 six-dimensional phase space, called the $\mu$ space. 
A system of $N$ particles is represented by a 
 point in a $6N$ dimensional phase space called $\Gamma$ space.}
 $(\vec{r}, \vec{p})$, and at time $t$.
Aim is to find an equation (of motion) for this function. 
The density changes with time since molecules enter and leave a
 given six  dimensional phase space volume element $d^3rd^3p$
at the phase space point ($\vec{r},\vec{p}$). 
Let $\vec{F}$ denote an external force
({\it e.g.} due to gravitation) acting on the molecules.
Suppose there are no collisions. A molecule at
$(\vec{r},\vec{p})$ at time $t$ will be found at 
$(\vec{r}+\vec{p}\ \Delta t/m, \vec{p} + \vec{F}\Delta t)$
at time $t+\Delta t$. Hamiltonian evolution preserves
volume element  $d^3rd^3p$ along a trajectory, called Liouville
theorem~\footnote{discovered by Joseph Liouville (1809-1882).}. 
Therefore,
\begin{eqnarray}
f(\vec{r}+\vec{p}\ \ \frac{\Delta t}{m},\ \vec{p}+\vec{F}\ \Delta t, 
t+\Delta t) 
&=& f(\vec{r},\vec{p},t)\ .
\end{eqnarray}
When there are collisions, we must add the contribution
from collisions and write~\footnote{Eq. (\ref{collision_term}) can be taken 
as definition of the collision term.}, 
\begin{eqnarray}\label{collision_term}
f(\vec{r}+\vec{p}\ \ \frac{\Delta t}{m},\ \vec{p}+\vec{F}\ \Delta t, t+\Delta t)
&=& f(\vec{r},\vec{p},t)
+\bigg(  \frac{\partial f}{\partial t}  \bigg)_{{\rm Col.}}
\Delta t\ .
\end{eqnarray}
Taylor-expanding  to first order in $\Delta t$ and taking the limit
$\Delta t\to 0$ we get,
\begin{eqnarray}
 \frac{\partial f}{\partial t}
=-\frac{1}{m}\vec{p}.\nabla_r f-
\vec{F}.\nabla_p f+
\bigg(  \frac{\partial f}{\partial t}  \bigg)_{{\rm Col.}},
\end{eqnarray}
where $\nabla_r$  and $\nabla_p$ are the gradient operators
with respect to position and momentum, respectively.
Boltzmann proposes a simple model for the collision term, see below.

Consider only binary collisions, true for a dilute  gas, 
where a pair of molecules with momenta $\vec{p}_1$ and $\vec{p}_2$ 
bounce off, after a collision, with momenta  $\vec{p^{\prime}}_{1}$ and 
$\vec{p^{\prime}}_{2}$,
respectively. Let $f(\vec{r},\vec{p}_1,\vec{p}_2,t)$ denote 
the density of pairs of particles with momenta  $\vec{p}_1$ and $\vec{p}_2$
at position $\vec{r}$ and at time $t$. 
\subsection{Stosszahlansatz} 
Boltzmann invokes stosszahlansatz -
collision number assumption - of Maxwell, which states,
\begin{eqnarray}
f(\vec{r}, \vec{p}_1,\vec{p}_2, t)=f(\vec{r}, \vec{p}_1, t)\  
f(\vec{r}, \vec{p}_2, t).
\end{eqnarray}
The above is also called the assumption of  molecular chaos. 
The momenta  of two particles are uncorrelated. The stosszahlansatz
is time symmetric. For both Maxwell  and 
Boltzmann, this assumption looked innocuous and  self evident.
From this,  Boltzmann derives an expression for the 
collision term, as described below. 

Let $d^3p_1, d^3p_2, d^3p^{\prime}_1$ and $d^3p^{\prime}_2$ be 
the momentum volume elements at $\vec{p}_1,\  \vec{p}_2,\  
\vec{p^{\prime}}_1,$ and $\vec{p^{\prime}}_2$ respectively.
Let us consider binary collisions that knock a molecule from
 $d^3p_1$ into $ d^3p^{\prime}_1$, while its collision partner 
gets knocked from $ d^3p_2$ into $d^3p^{\prime}_2$.
Since we are interested only in the collision term, we shall omit,
for notational convenience, reference to the dependence 
on position $\vec{r}$  and time $t$.
The rate at which these collisions take place is given by,
$$f(\vec{p}_1)d^3p_1 f(\vec{p}_2)d^3p_2\ \Sigma(\vec{p}_1,\vec{p}_2,
\vec{p^{\prime}}_1,\vec{p^{\prime}}_2)d^3 p^{\prime}_1d^3p^{\prime}_2 .$$
In the above, $\Sigma$ denotes the rate of transition from
$(\vec{p}_1,\vec{p}_2)$ to 
$(\vec{p^{\prime}}_1, \vec{p^{\prime}_2})$.  
The total rate of binary collisions that result in molecules
getting knocked out of volume element $d^3p_1$ is given by,
\begin{eqnarray}
R( {\rm OUT}) = d^3p_1 f(\vec{p}_1)\int d^3p_2\int d^3p^{\prime}_1
\int d^3p^{\prime}_2 \ \ f(\vec{p}_2)\ 
\Sigma(\vec{p}_1,\vec{p}_2,\vec{p^{\prime}}_1,\vec{p^{\prime}}_2).
\end{eqnarray}
While carrying out the integrals in the above, we must
ensure that momentum
 and 
energy  are conserved.
Let $R({\rm IN})$ denote the rate of binary collisions 
that knock molecules into the 
volume element
$d^3p_1$. This can be obtained  exactly the same way described above
except that we interchange the labels of
momenta before and after collision:  
$\vec{p}_1\leftrightarrow\vec{p^{\prime}}_1$ and 
$\vec{p}_2\leftrightarrow\vec{p^{\prime}}_2$.
In other words we consider binary collisions that knock molecules from 
$d^3p^{\prime}_1$ into $d^3p_1$ and from $d^3p^{\prime}_2$ into $d^3p_2$. 
We get,
\begin{eqnarray}
R( {\rm IN}) = d^3p_1\int d^3p_2\int d^3p^{\prime}_1\int d^3p^{\prime}_2\ \ 
f(\vec{p^{\prime}}_1)f(\vec{p^{\prime}}_2)\ 
\Sigma(\vec{p_{\prime}}, \vec{p^{\prime}}_2,\vec{p}_1,\vec{p}_2).
\end{eqnarray}
We consider  molecule - molecule interaction potential
to be spherically symmetric.  
We first note that a binary collision
is time 
symmetric
. In other words, the process seen in reverse is also an acceptable 
collision process. 
Hence, 
\begin{eqnarray}
\Sigma (\vec{p}_1,\vec{p}_2,\vec{p^{\prime}}_1,\vec{p^{\prime}}_2) =
\Sigma (-\vec{p^{\prime}}_1, -\vec{p^{\prime}}_2 , -\vec{p}_1,-\vec{p}_2).
\end{eqnarray}
Also $\Sigma$ is unchanged under simultaneous reflection  of all momenta:
\begin{eqnarray}
\Sigma (\vec{p}_1,\vec{p}_2,\vec{p^{\prime}}_1,\vec{p^{\prime}}_2) =
\Sigma (- \vec{p}_1, - \vec{p}_2, -\vec{p^{\prime}}_1, -  
\vec{p^{\prime}}_2).
\end{eqnarray}
Combining the above two we get,
\begin{eqnarray}
\Sigma (\vec{p}_1,\vec{p}_2,\vec{p^{\prime}}_1,\vec{p^{\prime}}_2) =
\Sigma (\vec{p^{\prime}}_1, \vec{p^{\prime}}_2 ,\vec{p}_1,\vec{p}_2).
\end{eqnarray}
Thus we can write the collision term as, 
\begin{eqnarray}\label{collision_contribution}
 \bigg(  
\frac{\partial f}{\partial t} \bigg)_{{\rm Col.}}
 =
\int\negthinspace d^3p_2\negthinspace\int\negthinspace
 d^3p^{\prime}_1 \negthinspace\int\negthinspace d^3p^{\prime}_2
\ \ \Sigma (  \vec{p}_1, \vec{p}_2 , 
\vec{p^{\prime}}_1, \vec{p^{\prime}}_2)  
 \bigg[ f(\vec{p^{\prime}}_1) f(\vec{p^{\prime}}_2) 
-  f(\vec{p}_1) f(\vec{p}_2)\bigg]\ . 
\end{eqnarray}
$\Sigma$ depends on the geometry of collision, the relative velocity 
of the two particles entering collision and the
 nature of the colliding particles.
The full  nonlinear Boltzmann transport equation  reads as,
\begin{eqnarray}\label{Boltzmann_equation}
 \frac{\partial f}{\partial t}\negthickspace\negthickspace
\negthickspace\negthickspace
& &  = -\frac{1}{m}\vec{p}_1.\nabla_r f-
\vec{F}.\nabla_p f+\nonumber\\
& &  \int d^3p_2\int d^3p^{\prime}_1\int d^3p^{\prime}_2
\ \ \Sigma (  \vec{p}_1, \vec{p}_2, 
\vec{p^{\prime}}_1, \vec{p^{\prime}}_2) 
 \bigg[ f(\vec{p^{\prime}}_1) f(\vec{p^{\prime}}_2)
-  f(\vec{p}_1) f(\vec{p}_2)\bigg] 
\end{eqnarray}
\section{Boltzmann H function}
 Boltzmann 
defines his famous $H$ function,
\begin{eqnarray}\label{H-function}
H(t)=\int d^3p\ \ f(\vec{p},t)\log\left[ f(\vec{p},t)\right],
\end{eqnarray}
and then shows that a density $f(\vec{p}, t)$
that solves the transport equation  
obeys,
\begin{eqnarray}
\frac{dH}{dt} \le 0.
\end{eqnarray}
 The above is clearly
time asymmetric. In contrast to Newtonian dynamics which does not 
distinguish the future from the past, the $H$-function has a well defined
direction of time, which is what the Second law is all about.
To prove the $H$ theorem, we write 
from Eq. (\ref{H-function}),
\begin{eqnarray}\label{dH_by_dt}
\frac{dH}{dt} = \int d^3p \left[ 1 + 
\log (f)\right]  \frac{\partial f}{\partial t} .
\end{eqnarray}
Therefore,
\begin{eqnarray}
\frac{\partial f}{\partial t} = 0 \ \ {\rm implies} \ \ \frac{dH}{dt}=0.
\end{eqnarray}
The $H$ function does not change with time when the system is in equilibrium.
Eq. (\ref{dH_by_dt}) in conjunction with the transport equation,
see Eq. (\ref{Boltzmann_equation}) yields after a few simple 
steps, the following expression for the time evolution of the 
$H$ function.
\begin{eqnarray}
\frac{dH}{dt} = -\frac{1}{4}\int d^3p_1\int d^2p_2 
\int d^3p^{\prime}_1 \int d^3p^{\prime}_2 
& &
\negthickspace\negthickspace\negthickspace\negthickspace
 \Sigma(\vec{p}_1,\vec{p_2},\vec{p^{\prime}}_1,\vec{p^{\prime}}_2)
\bigg[ f(\vec{p}_1)f(\vec{p}_2)-f(\vec{p^{\prime}}_1)f(\vec{p^{\prime}}_2)\bigg]
\times\nonumber\\
& & \bigg[ \log [  f(\vec{p}_1)f(\vec{p}_2)] -
\log[ f(\vec{p^{\prime}}_1)f(\vec{p^{\prime}}_2)\bigg]
\end{eqnarray}
We recognize that due to the concavity of the logarithm function,
\begin{eqnarray}
(y-x)(\log y -  \log x )\ge 0\ \ \forall\ \ x,y > 0
\end{eqnarray}
$H$ decreases with time monotonically giving rise to an arrow 
of time for macroscopic evolution.  
Thus Boltzmann, like a magician, produced
a time asymmetric rabbit from a time symmetric hat! 

The crucial  point  overlooked was in the usage of the stosszahlansatz
before and after collision. Momentum conservation,
$\vec{p}_1+\vec{p}_2=\vec{p^{\prime}}_1+\vec{p^{\prime}}_2$,
tells us that writing, 
\begin{eqnarray}
f(\vec{p^{\prime}}_1,\vec{p^{\prime}}_2) =
f(\vec{p^{\prime}}_1)\times f(\vec{p^{\prime}}_2)\ ,
\end{eqnarray}
is not correct, since a  pair of uncorrelated particles 
gets correlated after collision.
The {\it reversibility paradox} \cite{Loschmidt} 
of 
Josef Loschmidt (1821 - 1895)  
 and 
the {\it recurrence paradox} \cite{Zermelo} of
Ernst Zermelo~(1871~-~1956) 
 showed  Boltzmann's claim 
was untenable. Let me quickly tell what these two paradoxes are.  
\subsection{Loschmidt and reversibility paradox}
Loschmidt's argument was based on microscopic reversibility. Consider 
an isolated system that evolves from time $t=0$ to time $t=\tau$. 
Let there be a spontaneous increase of entropy during this evolution.
At time $t=\tau$ reverse the momenta of all the molecules. Allow the 
system to evolve from time $t=\tau$ to time $t=2\tau$. 
At time $t=2\tau$ reverse once again 
the momenta of all the molecules. Since the system obeys time-reversal
invariant Newtonian dynamics, it will end up at the same phase space 
point it started from. There would be a decrease in entropy during the 
evolution from time $t=\tau$ to time $t=2\tau$, contrary to the claim 
made by Boltzmann. This is called Loschmidt's reversibility paradox~\footnote{
Time reversal as discussed in the text can be  
implemented in a computer 
employing molecular dynamics simulation techniques. 
We find that even small errors in the calculations of positions and momenta
of the molecules are sufficient to reduce and eventually eliminate this effect.
The phase space trajectory of the macroscopic system is extremely 
unstable with respect to initial conditions. 
Two arbitrarily close trajectories move arbitrarily far apart
asymptotically. This is called chaos. This was known to 
Julius Henry Poincar\'e (1854 - 1912) \cite{Poincare_1890,
Poincare_1893}, 
a contemporary of Ludwig Boltzmann.
Chaos contains the seed for modern developments in 
non-equilibrium statistical mechanics.
We shall see more on these issues later.}.
\subsection{Zermelo and Recurrence paradox} 
Zermelo argued that an isolated system, 
under Hamiltonian dynamics will return 
arbitrarily close to its initial point in the 
phase space and infinitely often. This is called 
recurrence theorem,  discovered by Poincar\'e
\cite{Poincare_1890,Poincare_1893}. According to Poincar\'e recurrence
theorem, every dynamical
system is at least quasi-periodic if not exactly periodic. This follows from
Liouville theorem: a phase space volume of initial conditions evolve without 
change of its volume. Hence it is described by a tube shaped region of 
ever-increasing length. As the total region of phase space available 
to the dynamical system is finite, the tube must somewhere intersect itself.
This means that the initial and final states eventually come close to
each other. The dynamical system returns arbitrarily close to its initial 
state and it does so infinitely often.    
If there is a spontaneous increase of entropy 
during an interval of time,
there will be a spontaneous decrease of entropy 
during the interval of 
Poincar\'e recurrence; 
this contradicts 
 Boltzmann's claim~\footnote{Poincar\'e 
recurrence is easily
observed in systems with a very few degrees of freedom. 
But the recurrence time increases exponentially with the system size 
{\it i.e.} with the number of molecules. Hence Poincar\'e
recurrence is seldom observed in the thermodynamic limit.}.
\section{Statistical Entropy of Boltzmann}
Boltzmann  conceded
that perhaps, the use of stosszahlansatz has smuggled in an element 
of stochasticity (albeit in a very subtle way) into his otherwise
purely  dynamical derivation of the  transport equation. 
He contended correctly that his H theorem is violated only when the system
starts off from some special microstates which are  very small
in number. For an overwhelmingly large number of 
initial conditions, the dynamical evolution does obey 
the $H$ theorem.  In other words, the typical behaviour of a 
macroscopic system is
invariably consistent with the H theorem.  

Nevertheless,
in the year 1877, Boltzmann changed tack completely and 
 proposed a fully stochastic approach to the problem of 
macroscopic irreversibility.
He presented his
ideas in  a paper \cite{Boltzmann_1877}  
{\it on the relation between the Second law of 
thermodynamics and probability theory with respect to the law of thermal
equilibrium}. Of course 
Boltzmann interprets probability in a dynamical way: 
The probability of finding
a system in a region of its phase space is the fraction of 
the observation time the dynamical trajectory spends in that region.

Consider an isolated macroscopic system of $N$ particles. It is represented
by a point in a $6N$ dimensional phase space ($\Gamma$ space), 
moving incessantly along a
trajectory dictated by its dynamics. Let us coarse-grain 
the phase space in terms
of hyper cubes each of volume $h^{3N}$.  
It is like a 
graph sheet that coarse-grains a plane 
in terms of tiny squares. Here $h$ represents a constant having the
dimension of action~\footnote{Now we identify $h$ with 
Planck's constant; $h=6.626\times 10^{-34}$ Joules-second.}.
A phase space hyper cube is called a microstate.
Let $\vec{x}$ be the $6N$ dimensional vector denoting the phase space point
of the system and let $\rho(\vec{x},t)d^{6N}x$ be the probability of 
finding the system in an infinitesimal volume $d^{6N}x$ at $\vec{x}$ at time
$t$. Let the system be in equilibrium. In other words the density $\rho$ 
is independent of time. 
Let $\{ \rho_i\}$ denote the discrete representation of the phase space 
density $\rho(\vec{x})$. 

Boltzmann's $H$ function, see Eq. (\ref{H-function}), 
is then given by,
\begin{eqnarray}
H=\sum_{i=1}^{\hat{\Omega}} \rho_i\log (\rho_i)
\end{eqnarray}
where $\hat{\Omega}$ is the total number of microstates 
accessible to the system under macroscopic constraints of 
energy $U$, volume $V$ and number of molecules $N$. Boltzmann
defines entropy as,
\begin{eqnarray}\label{Gibbs_entropy}
S(U,V,N)=-k_B\sum_{i=1}^{\hat{\Omega}} \rho_i\log (\rho_i)\ ,
\end{eqnarray}
where $k_B$ is now called the Boltzmann 
constant~\footnote{$k_B=1.381\times 10^{-23}$ Joules per degree kelvin.}. 

If we assume that all the microstates are equally probable, then 
$\rho_i = 1/\hat{\Omega}\ \forall\  i$, and we get the famous formula 
for Boltzmann entropy,
\begin{eqnarray}\label{Boltzmann_entropy}
S=k_B\log (\hat{\Omega}), 
\end{eqnarray}
engraved on his tomb in Zentralfriedhof,
 Vienna~\footnote{Strangely, Boltzmann never
wrote down this formula  in any of his papers,
though he implied it. It was Max Planck who wrote
it down explicitly from the $H$ function.}. 
Notice  Boltzmann defines absolute entropy. In thermodynamics
only change in entropy is 
defined. 
\subsection{Is Boltzmann entropy consistent with thermodynamic entropy?}

Let $V$ be the number of coarse-grained volume cells occupied by
$N$ non interacting molecules. For simplicity we ignore the momentum
coordinates. Number of ways of 
configuring $N$ molecules in $V$ cells
is given by $\hat{\Omega}=V^N$, 
from which it follows~\footnote{This expression for entropy
is not extensive - called Gibbs' paradox. Boltzmann resolved 
the paradox by
introducing the notion of indistinguishable particles and 
corrected for over counting of microstates by dividing $\hat{\Omega}$ by
$N!$.\label{fn:Gibbs_paradox}}  $S=k_B N\log (V)$. 
Pressure is temperature times the partial derivative of 
entropy with respect to volume. We have,
\begin{eqnarray}
\frac{\partial S}{\partial V} = \frac{Nk_B}{V} = \frac{P}{T},
\end{eqnarray}
from which we get the ideal gas law: $PV=Nk_B T$. This leads to
\begin{eqnarray}
dS=\frac{\partial S}{\partial V} dV = \frac{1}{T}PdV.
\end{eqnarray}
Consider a quasi-static process in which the system draws 
a quantity $\dbar Q$ of reversible heat  and 
produces work equal to $PdV$. Thus $\dbar Q = P dV$, from 
which it follows that $\dbar Q = T dS$. 
Thus Boltzmann 
entropy is consistent with the thermodynamic entropy~\footnote{
The full expression for $\widehat{\Omega}(E,V,N)$ obtained 
taking into account the
momentum coordinates of the ideal gas molecules is given by the 
Sackur-Tetrode Equation, see below.
\subsection*{Sackur-Tetrode Equation}
\vglue -5mm
\begin{eqnarray}
\widehat{\Omega}(E,V,N)=\frac{1}{h^{3N}}\frac{V^N}{N!}
\frac{(2\pi mE)^{3N/2}}{\Gamma(\frac{3N}{2}+1)},\nonumber
\end{eqnarray}
where $E$ is the total energy of the isolated system, 
$m$ is the mass of a molecule, $h$ is Planck constant employed 
for coarse-graining the phase space ($h^{3N}$ is the volume 
of a $6N$ dimensional cube in units of which the phase space volume is 
measured)
and $\Gamma (\cdot)$ 
is the usual Gamma function,
\begin{eqnarray}
\Gamma (n) = \int_0 ^{\infty} dt\ \  t^{n-1}\ \ e^{-t}.\nonumber
\end{eqnarray}
The entropy of an ideal gas is thus given by,
\begin{eqnarray}
S(E,V,N)=Nk_B\log\bigg( \frac{E^{3/2}V}{N^{5/2}}\bigg)  + \frac{5Nk_B}{2} +
\frac{3Nk_B}{2}\log\bigg( \frac{4\pi m}{3h^2}\bigg)
.\nonumber
\end{eqnarray}
The above is known as Sackur-Tetrode equation.\label{Sackur_Tetrode}
}.
But Boltzmann  liberated  
entropy from its
thermal confines. We can now define entropy for a coin toss,
$S=k_B\log 2$ or  throw of a dice, $S=k_B\log 6$, {\it etc.}
In general 
if an experiment has $\hat{\Omega}$ outcomes and they are all
equally probable,
then we can associate an entropy, $k_B\log \hat{\Omega}$,
with the experiment.
\section{Boltzmann Entropy and Gibbs Entropy}
Consider an experiment of tossing $N$ identical and fair coins.
An outcome $\omega$ of this experiment is a string of Heads and Tails. We call 
$\omega$ a microstate. The set of all possible microstates of the experiment
is denoted by by $\Omega(N)$ called the sample space. The number of elements of the sample space is 
given by $\widehat{\Omega}(N)=2^N$. Let us count the number of Heads in a 
string $\omega$ and call it $n(\omega)$. 
The random variable $n$ can take any  value 
between $0$ and $N$. We call $n$ a macro state. Let $\Omega(n;N)=\{\omega:n(\omega)=n\}$ be the set of all 
strings having $n$ Heads. In other words it is  a set of all 
microstates belonging to the macro state $n$. The number of elements of the 
set $\Omega(n;N)$  or equivalently 
the number of microstates associated with the 
the macro state $n$, is given by
\begin{eqnarray}
\widehat{\Omega}(n;N)=\frac{N!}{n! (N-n)!}
\end{eqnarray}
We have 
\begin{eqnarray}
\widehat{\Omega}(N)=\sum_{n=0}^{N}\widehat{\Omega}(n;N)=2^N.
\end{eqnarray}
Boltzmann associates
an entropy $S$ with each macro state $n$: 
\begin{eqnarray}
S(n)=\log \bigg[ \widehat{\Omega}(n;N)\bigg].
\end{eqnarray}
 Note
we have set the Boltzmann constant $k_B$ to unity. Boltzmann postulates that 
the system, switching all the time from one microstate to 
another microstate, would evolve
in an entropy increasing way and eventually reach 
an equilibrium state characterized by 
an unchanging value of $n$ for which entropy is maximum. 
We immediately see that 
$\widehat{\Omega}(n;N)$ is maximum when $n=N/2$. 
Therefore Boltzmann entropy for the equilibrium system  is given by
\begin{eqnarray}\label{Boltzmann_entropy_coin_tossing}
S_B= \log\bigg[ \widehat{\Omega}(n=N/2;N)\bigg]
=\log\bigg[ \frac{N!}{ \left(\frac{N}{2}\right)!\ \ \left(\frac{N}{2}\right)!}\bigg]
\end{eqnarray}
 
Josiah Gibbs 
(1839 - 1903), proposed that equilibrium value of the macroscopic property
$n$ should be calculated by averaging over an appropriate  ensemble
of microstates. In the coin-tossing example considered here the 
ensemble consists of microstates from a Binomial distribution. 
Thus in Gibbs picture 
of statistical mechanics, 
\begin{eqnarray}
n_{{\rm eq}} = \langle n\rangle = 
\sum_{n=0}^{N} n\ \widehat{\Omega}(n;N)\frac{1}{2^N}=\frac{N}{2},
\end{eqnarray}
which is the same as that given by Boltzmann. 
Gibbs'
entropy, given by 
\begin{eqnarray}
S_G = N\log (2),
\end{eqnarray}
is different from Boltzmann's
entropy, see Eq. (\ref{Boltzmann_entropy_coin_tossing}).
  However, in the thermodynamic limit, 
Gibbs entropy and
Boltzmann entropy coincide. We have, in the limit of $N\to\infty$,
\begin{eqnarray}
 S_G  = S_B + {\cal O}(\log N).
\end{eqnarray}
\section{Gibbs Ensembles}
Gibbs developed statistical mechanics into  a fine tool for calculating  
equilibrium  properties of macroscopic systems
 as averages over  what we now call  
Gibbs' ensembles\cite{Gibbs_1902}.
\subsection{Microcanonical Ensemble}
The properties of an isolated system can be obtained 
by averaging over a microcanonical ensemble in which all microstates
are of the  same energy and occur with  the same probability. 
For example footnote (\ref{Sackur_Tetrode} ) expresses 
the number of microstates, $\widehat{\Omega}$
of $N$ ideal gas molecules confined to a volume $V$ and with energy $E$.

\subsection{Canonical Ensemble}
A closed system is one which exchanges only energy with the outside
world and not material or volume. It is described by a canonical ensemble.
The probability that a closed system will be found in a microstate
${\cal C}$,  
is  given by
\begin{eqnarray}
P({\cal C}) = \frac{1}{Z(T,V,N)}
\exp \bigg[ -\beta E( {\cal C})\bigg],
\end{eqnarray}
where  $\beta=1/(k_B T)$ and $Z(T,V.N)$ is called the canonical partition function given by
\begin{eqnarray}\label{canonical_partition_function}
Z(T,V.N) = \sum_{{\cal C}} \exp \bigg[ -\beta E( {\cal C})\bigg].
\end{eqnarray}

Let $\widehat{\Omega}(E,V,N)dE$ be the number of microstates
of the closed system having energy between $E$ and $E+dE$.  
We call  $\widehat{\Omega}(E,V,N)$ the density of states. 
We can express the canonical partition function as,
\begin{eqnarray}
Z(T,V,N)=\int dE \ \widehat{\Omega}(E,V,N)\ \exp(-\beta E)
\end{eqnarray}
The density of states is a rapidly increasing function of energy:
understandably so, since more the energy more is the number of ways 
of distributing it and hence more is the entropy. The exponential
function decreases with increase of energy. The 
product of these two will be a sharply peaked function 
energy, peaking at the thermodynamic energy $U=\langle E\rangle$.     
A saddle point estimate of the integral can be made and we get,
\begin{eqnarray}
Z(T,V,N)=\exp\bigg[ \frac{S}{k_B}-\beta U\bigg]
\end{eqnarray}
from which we get 
\begin{eqnarray}
F(T,V,N)=-k_B T\log Z(T,V,N) = U(S,V,N)-T S(U,V,N)
\end{eqnarray}
where $F(T,V,N)$ is the Helmholtz free energy, proposed
by Hermann von Helmholtz (1821-1894). 
\subsection{Grand canonical ensemble}
An open system is one which exchanges both energy and material 
with the outside world. It 
is described by a  grand canonical 
ensemble and the partition function 
is given by,
\begin{eqnarray}
{\cal Q}(T,V,\mu)= \sum_{{\cal C}}  \exp \bigg[ -\beta E( {\cal C})+\mu\beta N({\cal C})\bigg],
\end{eqnarray}
where $N({\cal C})$ is the number of molecules in the microstate ${\cal C}$
of the open system and $\mu$ is the chemical potential. 

We can construct different Gibbs ensembles 
depending on the system we are investigating. Gibbs provided 
a general framework  of statistical mechanics 
based on static Gibbs ensembles and averages over them. This is in contrast to
the ensemble of Boltzmann which is dynamical. It is the typical behaviour 
that forms  the basis  of Boltzmann's picture of 
statistical mechanics.

The  expression for entropy  given 
by Eq.  (\ref{Gibbs_entropy}) was also  derived by 
Claude Elwood Shannon (1916 - 2001),
in the context of information theory\cite{Shannon}. 
We say Eq. (\ref{Gibbs_entropy}) defines
Boltzmann-Gibbs-Shannon entropy~\footnote{Ever since, there have been several 
\lq entropies\rq\  proposed in different contexts. 
These include for example,
Fisher information \cite{Fisher_entropy},
von Neumann entropy \cite{Neumann_entropy}
 Renyi entropy\cite{Renyi},
Kolmogrov-Sinai entropy\cite{K-S_entropy}, 
Algorithmic entropy\cite{algorithmic_entropy},
Tsallis entropy\cite{Tsallis_entropy}
and Kaniadakis entropy \cite{Kaniadakis_entropy}.}.

Boltzmann entropy,
laid the foundation for statistical mechanics - 
a subject that helps us calculate 
macroscopic properties of an equilibrium  system from those of its
microscopic constituents and their interactions.
This subject has since grown to a very high degree 
of sophistication. More importantly the predictions of statistical 
mechanics  have been borne out by experiments.
Statistical mechanics   has become such a successful enterprise that 
physicists do  not anymore  question the use of statistics
for describing macroscopic phenomena~\footnote{Physicists were 
in for a greater embarrassment
with the advent of quantum mechanics. 
Statistics enters into microscopic laws. Stochasticity is intrinsic 
to quantum mechanics. The notion of ensemble of Maxwell, Boltzmann and 
Gibbs came in handy in describing the results of measurements in 
quantum mechanics.  
}.
But the nagging doubt remains:
What is the origin for the observed stochasticity ?
\section{Dynamical Entropy from  Chaos}
Then came a meteorologist and mathematician named  Edward Norton Lorenz 
with his three coupled first-order nonlinear differential equations. 
He had obtained them
by truncating Navier-Stokes
equations~\footnote{The Navier-Stokes equation of Claude Louis Marie Navier (1785-1836) 
and  Sir George Gabriel Stokes (1819-1903), 
is the primary equation of computational fluid dynamics, 
relating pressure and external forces acting on a fluid to the response of the 
fluid flow. Forms of this equation are used in computations for 
aircraft and ship design, weather prediction, and climate modeling.  
}.
The three equations of Lorenz were intended to provide a 
simple and approximate description of atmospheric behaviour. 
Lorenz was solving them on a computer.
He discovered \cite{Lorenz_1963} 
that he had two very different numerical solutions for the 
same problem with almost identical initial conditions. 
This chance observation heralded 
a new field called chaotic dynamics \cite{Chaos_books,Ruelle_1989}. 
Two phase space trajectories of a chaotic system starting off from arbitrarily
close 
phase space points diverge exponentially and become completely uncorrelated
asymptotically.
This means that you can not have any hope of making any long term
predictions from deterministic
equations if they happen to be chaotic. 
In other words {\it determinism does not necessarily
imply predictability}. 

Possibility of dynamical instability due to sensitive dependence on initial
conditions was known to 
Poincar\'e \cite{Poincare_1890,Poincare_1893}. I have 
already mentioned of this while discussing Loschmidt's 
reversibility paradox \cite{Loschmidt} 
and Zermelo's recurrence paradox \cite{Zermelo}. 
But the full import of  Poincar\'e's findings
was lost on the physicists for over half a century.  
They  did not think much of it until computers arrived on their
desktops and helped them see on graphic terminals, 
the  strange trajectories traced 
by chaotic dynamical systems. 
 
A standard way of determining whether or not a dynamical system is
chaotic is to calculate the Lyapunov exponent. 
There are as many Lyapunov exponents 
as the dimensions of the
phase space. Consider dynamics in an $n$ dimensional phase space.
Consider an $n$ - dimensional sphere of initial conditions.
At  a later time all the trajectories emanating from the sphere
will form an $n$-dimensional ellipsoid. We can calculate a Lyapunov exponent
for each dimension. When talking of a single Lyapunov 
exponent we normally refer to the largest and denote it by 
$\lambda$.  Thus if  $\lambda > 0$, 
we say the 
dynamics is  chaotic. 
The system becomes unpredictable for times 
greater than of the order of $1/\lambda$. 
On these asymptotic time scales the 
system becomes amenable to statistical description. 

We recognize thus, that at least in principle,  Chaos 
provides {\it raisin d'etre} for statistics in statistical mechanics.
All systems that obey the laws of thermodynamics are chaotic.
Nonlinear dynamics and  chaos provide 
the link between 
deterministic micro and the stochastic macro -  a link that Boltzmann
was struggling to figure out.

In fact Boltzmann's  interpretation of  
probability was entirely based on dynamics.   
The dynamical trajectory of an isolated equilibrium system is confined 
to  a constant energy  surface in  a $6N$ 
dimensional phase space.  
Boltzmann first shows that the phase space density $\rho$ remains constant
along a trajectory;
this is now called the Liouville theorem.  
He then assumes that all the points on the energy surface 
lie on a single trajectory.
This is called ergodicity.  
Then $\rho (\vec{x}) =\delta ( {\cal H}(\vec{x}) - E)$ is the 
stationary density, where  
$\cal{H}$ is the Hamiltonian,  $E$ is energy and $\delta$ is the 
usual Dirac delta function.

Boltzmann's  ergodicity has been generalized by 
Sinai \cite{Sinai_1968}, Ruelle \cite{Ruelle_1980} 
and Bowen \cite{Bowen_1970} to describe dissipative systems in a 
steady state. The strange  attractor of the dissipative 
dynamics is the non-equilibrium analogue of the equilibrium constant energy
surface considered by Boltzmann.  
The SRB measure
\cite{Ruelle_1980,ER_1985} on the attractor expressed 
in terms of phase space volume contraction 
is analogous to the Liouville measure on the energy surface
of an equilibrium isolated system. Such a generalization permits 
assignment of dynamical weights to  non-equilibrium states. 
These weights, let me repeat, are based on the dynamical properties 
of the microscopic constituents of a  macroscopic system. To appreciate
the import of this statement, we must recognize that  words like
equilibrium, heat, entropy, temperature {\it etc.}, belong to the 
vocabulary of the  macroscopic world of thermodynamics. 
They do not have any meaning in the 
microscopic world
.
Paraphrasing Maxwell,
{\it at microscopic level you can not tell heat from work, since both are
essentially energy}, see below. 

\subsection{Microscopic description of work and heat}
\label{microscopic_heat_and_work}
Let $U$ denote the thermodynamic energy of a 
closed system obtained by averaging the statistical mechanical energy 
$E$, over a canonical ensemble of microstates. 
Let the microstates accessible to the system be indexed by 
natural numbers $i$. Let $p_i$ denote the probability for the system to be 
in microstate $i$ whose energy is $E_i$ 
. The thermodynamic energy $U$ is then given by, 
$$U=\sum_i p_i E_i\ .$$ We have formally,
\begin{eqnarray}
dU &=&  \sum_i
\frac{\partial U}{\partial E_i}dE_i
+
\sum_i \frac{\partial U}{\partial p_i}dp_i\nonumber\\
&= &\sum_i p_i d E_i +\sum_i E_i dp_i\ . 
\end{eqnarray}
Thus we can change the energy of a system by an amount $dU$,
  through 
work $\dbar W$, given by the first term on the 
right and/or heat $\dbar q_{{\rm rev}}$, given by
the second term on the right in the above equation.
 Thus work and heat are simply two
modes of energy transfer.
\subsubsection*{Work}
To identify the first term as work we proceed as follows. We have,
\begin{eqnarray}
\sum_i p_i d E_i &=&
\sum_i p_i \frac{\partial E_i}{\partial V} dV\nonumber\\
& & \nonumber\\
&=&\left( \frac{\partial }{\partial V}\sum_i p_i E_i\right) dV\nonumber\\
& & \nonumber\\
&=&\frac{\partial U}{\partial V} dV\nonumber\\
& & \nonumber\\
&=&-PdV\nonumber\\
& & \nonumber\\
&=&\dbar W\ ,
\end{eqnarray}
where $P$ denotes pressure. Work corresponds to change in energy of 
the macroscopic system brought about by changing the energies of its
microstates without altering in any way their probabilities $\{ p_i\}$. 
\subsubsection*{Heat}
To identify the second term as heat, we start with the definition of 
Boltzmann-Gibbs-Shanon entropy 
$$S=-k_B\sum_i p_i\log p_i ,$$ and proceed as follows.
\begin{eqnarray}
\dbar q_{{\rm rev}} &=& T dS\nonumber\\
&=&-k_B T\sum_i dp_i - k_B T\sum \log p_i dp_i\nonumber\\ 
& & \nonumber\\
&=&-k_B T\sum_i dp_i - k_B T\sum \log p_i dp_i\nonumber\\ 
& & \nonumber\\
&=&-k_B T\sum_i dp_i \log p_i\nonumber\\
& & \nonumber\\
& = & k_B T \sum_i dp_i \bigg[ \beta E_i + \log Z\bigg]\nonumber\\
& & \nonumber\\
&=&\sum_i   E_i \ dp_i\ ,
\end{eqnarray} 
where $Z$ denotes the canonical partition function,
see Eq. (\ref{canonical_partition_function}) 
. In the above derivation we have made use of the fact that
in a canonical ensemble describing a closed system, 
$p_i=Z^{-1}\exp (-\beta E_i)$. Thus heat is change of energy of a 
closed system brought about by changing the  probabilities $\{ p_i\}$ 
without altering in any way the energies $\{ E_i\}$ of the microstates. It is in fact 
because of this identification we relate  heat and hence 
entropy to randomness. 

In the phase space of the thermodynamic variables
only an equilibrium system can be represented by a point; only a 
quasi-static process can be represented by a curve.
However in the $6N$ dimensional phase space of 
statistical mechanics, a macroscopic system 
in equilibrium or not, can be represented by a point; any process
can be represented by a trajectory, to  which we can attach a 
suitably defined dynamical weight.

Thus dynamical measures of recent times,
have liberated the notion of entropy from its equilibrium 
and quasi-static confines, into 
non-equilibrium realms. We have, indeed, come a long way: 
from the thermal entropy of Clausius to 
the statistical entropy of Boltzmann (both
applicable to equilibrium systems and quasi-static processes), 
and now to the 
SRB measures (defined for non-equilibrium systems and processes). 
Recently SRB measure has been shown to provide a  correct 
description \cite{GC_1995} of a far from equilibrium system in a computer 
simulation \cite{ECM_1993}.
\section{Entropy Fluctuation Theorems}
These new developments are embodied in what we call
 fluctuation theorems \cite{ECM_1993,ft_literature}. 
The general idea behind a fluctuation  theorem can be 
stated as follows. 
Let $S_{\tau}$ denote  
 entropy production rate 
calculated by averaging over segments of a long 
trajectory of duration $\tau$. Note that $S_{\tau}$ is a dynamical
entropy obtained from observing the phase space expansion/contraction.
Let $\Pi(S_\tau)$ be the 
probability of $S_\tau$. This can be calculated by considering an ensemble
of long trajectories each of duration $\tau$. 
Fluctuation theorem states,
\begin{eqnarray}
\frac{\Pi(S_\tau)}{\Pi(-S_\tau)} = \exp [\tau  S_\tau].
\end{eqnarray}
Fluctuation theorem helps us calculate the probability for the entropy
to change in a way opposite to that dictated 
by the Second law; this probability 
of Second law violation is exponentially small for large systems and for  
long observation times.  By the same token fluctuation theorems 
predict and more importantly quantify  
Second law violation 
in  small systems and  on small time scales of observation.
The predictions of fluctuation  theorems  have since been 
verified experimentally \cite{CRWSSE,WSMSE}. See also \cite{ONAD}
for an interesting examination of the experimental tools of 
fluctuation theorems. 
\section{Jarzynski Identity}
In the year 1997, C. Jarzynski\cite{Jarzynski}
discovered a remarkable identity relating non-equilibrium
work fluctuation to equilibrium free energies. Consider a switching process,
discussed earlier, carried out over a time $\tau$, 
with the system  
thermostatted~\footnote{A thermostat  
exchanges energy with the system
without changing its temperature or performing any work}
at temperature $T=1/(k_B\beta)$. 
Let $W$ denote the work done during the switching
process.
We carry out the switching several times and collect an ensemble  $\{ W_i\}$,
formally represented by the probability density $\rho(W;\tau)$.
All the switching experiments are carried out with the same  
protocol. 
 If $\tau=\infty$, the process is quasi-static. We have $W_i=W_R\ \forall \ i$.
The work done is 
called reversible work, $W_R$. For a general switching experiment
where $\tau \ < \infty$,  the Second law says that 
\begin{eqnarray}
\Delta F \le \langle W\rangle
,
\end{eqnarray}
where $\Delta F$ is the change in the  Helmholtz free
energy
.

Jarzynski's identity  
is given by,
\begin{eqnarray}
\bigg\langle \exp (-\beta W)\bigg\rangle =\exp\left( -\beta \Delta F\right),
\end{eqnarray}
where $\langle \cdot\rangle$ denote averaging over the distribution of $W$.
\subsection{Jarzynski identity and the Second law}
It may be noticed that since the exponential function is
convex, we have,
\begin{eqnarray}
\bigg\langle \exp (-\beta W  )\bigg\rangle \ge 
\exp \bigg[ -\beta\left\langle W\right\rangle\bigg],
\end{eqnarray}
which in conjunction with Jarzynski's identity implies that,
\begin{eqnarray}
\exp (-\beta\Delta F ) & \ge & 
\exp \bigg[ -\beta\left\langle W\right\rangle\bigg] ,\nonumber\\
& & \nonumber\\
-\beta\Delta F   &    \ge  &  -\beta\langle W\rangle ,\nonumber\\
& & \nonumber\\
\Delta F & \le & \left\langle W\right\rangle ,
\end{eqnarray}
which is a statement of the Second law.
In this sense, proof of Jarzynski's identity is a proof of the
Second law.
\subsection{Jarzynski identity: cumulant expansion}
Let us express Jarzynski's equality as a cumulant expansion 
\cite{KPN_2004},
\begin{eqnarray}\label{cumulant_expansion}
\bigg\langle \exp (-\beta W)\bigg\rangle  \equiv
\exp\left[ \sum_{n=1}^{\infty} \frac{ (-\beta)^n \zeta_n}{n!}\right]
=\exp (-\beta \Delta F)\ ,
\end{eqnarray}
where $\zeta_n$ denotes the $n-$th cumulant of $W$. The cumulants and  the moments are
related to each other. 
The $n$-th cumulant can be expressed in terms of the
moments of order $n$ and less. The first cumulant, $\zeta_1$ is the same as 
the first moment $\langle W\rangle$; the second cumulant, 
$\zeta_2$ is the variance
$\sigma^2= \langle W^2\rangle -\langle W\rangle ^2$; {\it etc.}
From the cumulant expansion given by  Eq. (\ref{cumulant_expansion}),  we get,
\begin{eqnarray}\label{Jarzynski_cumulants}
\Delta F &=& \left\langle W\right\rangle -\frac{1}{2}\beta\sigma_W ^2 +
\sum_{n=3}^{\infty} {{(-\beta)^{n-1} \zeta_n}\over{n!}}\ .
\end{eqnarray}
\subsubsection{Reversible work and free energy}
Consider a quasi-static  switching process for which, 
\begin{eqnarray}
\rho~(W;\tau =\infty)=\delta~(W-W_R) \ ,
\end{eqnarray}
by definition. Then, in 
Eq. (\ref{Jarzynski_cumulants}), only the first term (of the cumulant
expansion) is non-zero. We get, 
\begin{eqnarray}
\left\langle W\right\rangle = W_R =  \Delta F\ ,
\end{eqnarray}
 consistent 
with thermodynamics.
\subsubsection{Fluctuation and dissipation }
Now consider a switching process, during which the system remains very close to
equilibrium; it is reasonable to expect the statistics of $W$ to obey
the Central Limit Theorem~\footnote{
According to the Central Limit
Theorem, the sum of $N$ independent and identically distributed, 
finite variance random variables, has an asymptotic ($N\to\infty$) 
Gaussian distribution with both  mean and variance  diverging  
linearly with $N$.  This means that the relative
fluctuation is inversely proportional to $\sqrt{N}$ and hence is small
for large $N$. 
See {\it e.g.} \cite{Papoulis}.
This is easily seen as follows. 
\subsection*{The Central Limit Theorem}
Let $$Y=\frac{1}{\sqrt{N}} \sum_{i=1}^{N}X_i\ ,$$ 
where $\{ X_i : i=1,N\}$ are identically distributed
independent random variables with zero mean and finite variance $\sigma^2$. Let $\Phi_X (k)$
denote the characteristic function of the random variable $X$. Then 
\begin{eqnarray}
\Phi_Y (k) &=&  \bigg[ \Phi_X (k\to\frac{k}{\sqrt{N}})\bigg] ^N\nonumber\\
&=&\exp \bigg[-\frac{1}{2}k^2\sigma^2
+\sum_{n=3}^{\infty} \frac{(ik)^n}{n!}N^{-(n-2)/2}\ \zeta_n\bigg]\nonumber\\
&{}^{\ \ \ \ \sim}_{N\to\infty}& \exp\bigg[ -\frac{1}{2}k^2\sigma^2 + {\cal O}(1/\sqrt{N})\bigg]
\nonumber
\end{eqnarray}
where $\zeta_n$ denotes the $n-$th cumulant of $X$. The Fourier inverse of 
the asymptotic ($N\to\infty$) expression for $\Phi_Y (k)$ is Gaussian with mean zero and variance 
$\sigma^2$. 
\label{clt}}.
Hence $\rho (W;\tau >> 0)$ shall be a Gaussian;
for a Gaussian, all the cumulants from the third up-wards 
are identically zero; hence, in Eq. (\ref{Jarzynski_cumulants}),  
only the first two terms survive and we get
\begin{eqnarray}
\Delta F =  \langle W\rangle -\frac{1}{2}\ \beta\ \sigma_W ^2  \ .
\end{eqnarray}
Dissipation given by, 
\begin{eqnarray}
\langle W_d \rangle =\langle W\rangle -\Delta F=\frac{1}{2}\beta\sigma_W ^2 ,  
\end{eqnarray}
is proportional to fluctuation, $\sigma^2_W$. 
This result is identical to the fluctuation  
dissipation relation of Callen and Welton \cite{Callen}.
See \cite{Van_den_Broeck} for an interesting discussion on 
Gaussian Work fluctuation, Jarzynski identity and 
fluctuation dissipation theorem.  
However, if  the switching process
drives the system far from equilibrium, the work distribution would  no longer
be  Gaussian and we need to include contributions from higher order cumulants to
calculate the dissipation $\langle W_d\rangle $  and hence free energy:
$\Delta F = \langle W\rangle - \langle W_d\rangle $.
Jarzynski's equality has been shown to hold good for 
Hamiltonian evolution\cite{Jarzynski} as well as stochastic
evolution\cite{Jarzynski_Stochastic}; its validity has been established 
in computer simulation\cite{Jarzynski_Stochastic} and in 
experiments\cite{Jarzynski_Expt}. 
\section{Microscopic Reversibility and Crooks identity}
In another parallel, independent and interesting development, 
Gavin E. Crooks \cite{Crooks} discovered a fluctuation theorem for  
a thermostatted,  Markovian dynamical process. During the process, 
the degree of freedom    
$\Lambda$ switches from an initial value of $\Lambda_0$ to a final value
$\Lambda_N$ in $N$ time steps. The switching process is not necessarily
quasi-static. 
\subsection{Heat Step,  Work step and Markov Chain}
The system is initially in a microstate 
C$_0(\Lambda_0)\in\Omega (\Lambda_0)$, 
where $\Omega(\Lambda_0)$ denote the set of all microstates of the 
system with $\Lambda = \Lambda_0$. 
Each step is considered as made up of a heat sub-step:
C$_0(\Lambda_0)\to {\rm C}_1 (\Lambda_0)$ 
and a work sub-step: C$_1 (\Lambda_0)\to {\rm C}_1 (\Lambda_1)$.
Thus we get a Markov
chain~\footnote{A Markov chain describes the time evolution of a 
system with a finite or countable number of microstates. 
We also specialize to Markov chain in discrete time. In general
for a Markov process, the past has no influence over the future once the 
present is specified.} 
of microstates given by,
\begin{eqnarray}
{\cal F}\vert {\rm C}_0 (\Lambda_0)=
{\rm C}_0(\Lambda_0) \to {\rm C}_1(\Lambda_0)\to {\rm C}_1(\Lambda_1)\to
\cdots \to {\rm C}_k(\Lambda_k)\to {\rm C}_{k+1}(\Lambda_k)\to
{\rm C}_{k+1}(\Lambda_{k+1})\to \nonumber\\
\cdots\to  {\rm C}_{N-1}(\Lambda_{N-1})
\to  {\rm C}_{N}(\Lambda_{N-1})\to {\rm C}_{N}(\Lambda_{N})\nonumber
\end{eqnarray}
 Let us consider a heat sub-step 
${\rm C}_k(\Lambda_k)\to {\rm C}_{k+1}(\Lambda_k)$, described by a 
Markov transition matrix $M(k)$ whose elements are given by
\begin{eqnarray}
M_{i,j}(k) = P({\rm C}_{k+1}={\cal C}_i\vert {\rm C}_{k}={\cal C}_j),
\end{eqnarray}
where ${\cal C}_i \in \Omega (\Lambda_k)$. We have used
script symbol ${\cal C}$ to denote microstates of the system and 
roman symbol ${\rm C}_k$ to denote those
on the Markov chain with $k$ serving as the time index. 
The matrix $M(k)$ has the following properties:
\begin{enumerate}
\item[$\bullet$]
The elements of $M(k)$ are all non-negative: $$M_{i,j}(k) \ge 0\ 
\forall\ i,j.$$
Note $M_{i,j}$ denotes (transition) probability.  
\item[$\bullet$]
$M(k)$ is column stochastic: $$\sum_i M_{i,j} (k)=1\ \forall \ j.$$
This follows from the normalization. After a step the system
must be  found in any one of its microstates with unit probability. 
\item[$\bullet$]
$M(k)$ is regular: There exists an integer $n > 0$, such that 
$$\bigg( \left[ M(k)\right] ^n\bigg)_{i,j}\  > \ 0
\ \ \forall\ \   i,j.$$ This ensures ergodicity.
\item[$\bullet$]
$M(k)$ is balanced: There exists a unique invariant probability
vector~
\footnote{Peron-Frobenius theorems, see {\it e.g.}  \cite{Gantmacher}, 
tell us the following.
The largest eigenvalue of $M$ is real and non degenerate. Its value is unity. 
All other eigenvalues of $M$ are much less than unity in modulus. The right
eigenvector
associated with the largest eigenvalue is called the invariant or equilibrium
probability vector and is denoted by $\vert\pi\rangle$. 
The eigenvectors of $M$ are linearly independent and span
the vector space of $M$. We have $M^n\vert\phi\rangle \to \vert\pi\rangle$,
for $n\to\infty$ and for $\langle \phi\vert \pi\rangle \ne 0$. 
Physically it means 
that the system eventually relaxes to its equilibrium state 
starting from any arbitrary non-equilibrium state. Also once the 
system reaches equilibrium it continues to be in equilibrium. Further action
of $M$ does not change its state. $M\vert\pi\rangle=\vert\pi\rangle$.}
$\vert \pi(k)\rangle$ such that
 $$M(k)\vert\pi(k)\rangle = \vert\pi(k)\rangle .$$
\item[$\bullet$]
$\vert\pi(k)\rangle  
$ describes the equilibrium distribution of the closed system at 
$\beta$ and with $\Lambda = \Lambda_k$. The components
of $\vert\pi\rangle$ are given by,
\begin{eqnarray}
\pi_i (k) =\frac{\exp\bigg[ -\beta 
E({\cal C}_i,\Lambda_k)\bigg]}{Z(\beta,\Lambda_k)},
\end{eqnarray}
where $E({\cal C}_i,\Lambda_k)$ is the energy  of the 
microstate ${\cal C}_i$ belonging to the system with $\lambda=\Lambda_k$. 
The canonical partition function is denoted by $Z(\beta,\Lambda_k)$. 
\end{enumerate}
\subsection{Metropolis and Heat-bath algorithms} 
We need a model for $M(k)$. For example
Metropolis algorithm\cite{Metropolis} prescribes, 
\begin{eqnarray}
M_{i,j}(k)&=&\alpha\times min\left( 1,\frac{\pi_i(k)}{\pi_j (k)}\right)
\ \ \  \forall\ i,j  \ \ \ {\rm and}
\ i\ne j ,\\
& & \nonumber\\
M_{i,i}(k) & = & 1-\sum_{j\ne i}M_{j,i}\ \forall \ i , 
\end{eqnarray}
where the constant $\alpha$ has been introduced to ensure that 
no diagonal element is negative or exceeds unity.

The heat-bath algorithm \cite{Cruetz} also known as 
Glauber algorithm \cite{Glauber}  or
Gibbs' sampler \cite{Gibbs_sampler} is given by 
\begin{eqnarray}
M_{i,j} = \frac{\pi_i}{\pi_i + \pi_j} 
\ \ \forall\ \  i,j
\ .
\end{eqnarray}

Once a model for $M$ is defined, we can calculate the probability
for the Markov chain ${\cal F}\vert {\rm C}_0(\Lambda_0)$, where
we take each work sub-step with unit probability.
\subsection{Time-Reversal of Markov Chain}
Let us now run the Markov chain back-wards and call it the time - reversal.
Let ${\cal R}\vert {\rm C}_N (\Lambda_N)$ denote the 
time reversal of ${\cal F}\vert {\rm C}_0(\Lambda_0)$. It is given by,
\begin{eqnarray}
{\cal R}\vert {\rm C}_N(\Lambda_N) &=&
 {\rm C}_{N}(\Lambda_{N})\to {\rm C}_{N}(\Lambda_{N-1})\to
 {\rm C}_{N-1}(\Lambda_{N-1})\to\cdots\nonumber\\
& & \to {\rm C}_{k+1}(\Lambda_{k+1})
\to{\rm C}_{k+1}(\Lambda_k)\to {\rm C}_k(\Lambda_k) 
\cdots\to{\rm C}_1(\Lambda_1)\to 
{\rm C}_1(\Lambda_0) \to {\rm C}_0(\Lambda_0)\nonumber
\end{eqnarray}
Note that in the time-reversed Markov chain,
the work sub-step comes first followed by the heat sub-step
in every time step.
We need to calculate the probability for the time reversed 
Markov chain. Reversing the work sub-step is easily visualized.
We switch the 
parameter $\Lambda$  back-wards with unit probability. 
Let the time reversed heat step 
${\rm C}_{k+1}(\Lambda_k)\to
{\rm C}_{k}(\Lambda_k)$ 
be described by $\widehat{M}(k)$,
called the  
reversal of $M(k)$. 
To construct $\widehat{M}(k)$ we proceed as follows.

 Consider the heat sub-step 
${\rm C}_{k+1}(\Lambda_k)\to
{\rm C}_{k}(\Lambda_k)$ 
in the forward Markov chain. Define a 
two-step  joint probability matrix
$W$  whose elements are given by,
\begin{eqnarray}
W_{i,j} (k) &=& P\bigg({\rm C}_{k+1}(\Lambda_k)={\cal C}_i,
{\rm C}_{k}(\Lambda_k)={\cal C}_j\bigg)\nonumber\\
& & \nonumber\\
&=&M_{i,j}(k)\pi_j(k)
\end{eqnarray}
In the above the second step follows from  the definition of conditional probability. 
Thus, given $M(k)$ we can get the corresponding $W(k)$ and {\it vice versa}.
To this end we define a diagonal matrix $D(k)$ with elements,
\begin{eqnarray}
D_{i,j}(k)=\pi_i (k)\delta_{i,j}.
\end{eqnarray}
Then,
\begin{eqnarray} 
W(k)=M(k)D(k)
\end{eqnarray}
and 
\begin{eqnarray}
M(k)=W(k)D^{-1}(k)
\end{eqnarray}
Also it is easily checked that 
$W(k)$ is matrix-stochastic: 
\begin{eqnarray}
\sum_i\sum_j W_{i,j}(k)=1.
\end{eqnarray}
Let $\widehat{W}(k)$ denote the time 
reversal of $W(k)$. A little thought will convince you that a 
good choice of $\widehat{W}(k)$ is $W^{\dagger}(k)$, where
the superscript ${}^{\dagger}$ denotes transpose operation.  
The corresponding
$\widehat{M}(k)$ can be obtained as follows.
\begin{eqnarray}
\widehat{M} (k)& = &\widehat{W}(k)D^{-1}(k)\nonumber\\
            & & \nonumber\\  
            & = & W^{\dagger}(k)D^{-1}(k)\nonumber\\
            & & \nonumber\\  
            & = & D(k)M^{\dagger}(k)D^{-1}(k). 
\end{eqnarray}
We say a Markov chain is time symmetric if $W$ is symmetric. 
In other words,
\begin{eqnarray}
\widehat{W}(k)=W^{\dagger}(k)=W(k),
\end{eqnarray}
for time symmetry. 
Also for a time symmetric
Markov chain, we have 
\begin{eqnarray}
\widehat{M}(k)&=&\widehat{W}(k)D^{-1}(k)\nonumber\\
            & & \nonumber\\  
           &=&W^{\dagger}(k)D^{-1}(k)\nonumber\\
            & & \nonumber\\  
           &=&W(k)D^{-1}(k)\nonumber\\
            & & \nonumber\\  
           &=&M(k).
\end{eqnarray} 
This implies that 
$\pi_j (k)M_{i,j} (k)= \pi_i (k)M_{j,i}(k)$, 
called detailed balance~\footnote{
The Metropolis  \cite{Metropolis} and the heat-bath algorithms \cite{
Cruetz,Glauber,Gibbs_sampler}  obey detailed balance.
There are algorithms that do not obey detailed balance.
It is often said that a simple balance condition, 
$M
\vert\pi\rangle=\vert\pi\rangle$ is adequate to drive the system
to
equilibrium in a computer simulation, see {\it e.g.} 
\cite{VIMMWD,ONAPY}.
We see that it is detailed balance that ensures time symmetry 
in a sequence of microstates visited by the system after equilibration.
If the computer algorithm obeys only balance 
and not detailed balance then time asymmetry 
in the Markov chain of microstates sampled, would be present even 
during equilibrium runs. 
}. 
A sequence of microstates visited by an  equilibrium 
system constitutes a time-symmetric Markov chain. 
\subsection{Crooks Identity}
The probability of ${\cal R}\vert {\rm C}_N (\Lambda_N)$ 
can be calculated from the matrices 
$\{ \widehat{M}(k)\}$. Let $\Pi_{{\cal F}}$ denote the 
probability of  ${\cal F}\vert {\rm C}_0(\Lambda_0)$ and 
$\Pi_{{\cal R}}$ that of its reverse. The ratio of these two probabilities
can be calculated and is given by,
\begin{eqnarray}
\frac{\Pi_{{\cal F}}}{\Pi_{{\cal R}}} = \exp \bigg[ -\beta Q({\cal F})\bigg],
\end{eqnarray}
where $Q$ is the energy absorbed by the system, in the form of heat, from the 
thermostat during forward Markov chain evolution.  
The above is called Crooks identity. The import of Crooks' finding 
can be understood if we consider  switching 
from an equilibrium ensemble at $\beta$ and 
with $\Lambda=\Lambda_0$ to another equilibrium ensemble at 
the same $\beta$ but with 
$\Lambda=\Lambda_N$ through a process which is not necessarily quasi-static. 
Thus ${\rm C}_0(\Lambda_0)$
and ${\rm C}_N(\Lambda_N)$ belong to equilibrium ensembles 
at the same temperature.
Then,
\begin{eqnarray}\label{Jarzynski_2}
\frac{\Pi\bigg( {\rm C}_0(\Lambda_0)\bigg)}
{ \Pi\bigg( {\rm C}_N(\Lambda_N)\bigg)}\times
\frac{\Pi_{{\cal F}}}{\Pi_{{\cal R}}} &=& 
\frac{ \exp\bigg[-\beta E\bigg( {\rm C}_0(\Lambda_0)\bigg)\bigg]}{Z(\beta,\Lambda_0)}
\frac{Z(\beta,\Lambda_N)}{ \exp\bigg[-\beta E\bigg( {\rm C}_N(\Lambda_N)\bigg)\bigg]}
\times \exp(-\beta Q)\nonumber\\
   & & \nonumber\\
   & & \nonumber\\
&=& \exp\bigg[ \beta\bigg\{ \Delta E -\Delta F -Q\bigg\} \bigg]\nonumber\\
    & & \nonumber\\
   & & \nonumber\\
&=& \exp\bigg[ \beta\bigg\{ \langle W\rangle -\Delta F\bigg\}\bigg]\nonumber\\
    & & \nonumber\\
   & & \nonumber\\
&=& \exp\bigg[\beta \langle W_d\rangle\bigg].
\end{eqnarray} 
In the above we have used the definition of free energy,
$F(T,V,N)=-k_B T\log Z(T,V,N)$ for going from the 
first line to the second line. In going to the third line
from the second, we have made use of the first law: $\langle W\rangle =
\Delta E-Q$. 
 Physically Eq. (\ref{Jarzynski_2}) 
 means that the probability of finding a dissipating segment of 
a Markov chain evolution is exponentially large compared to 
that of finding its reverse. Starting from Crooks identity we can 
derive fluctuation theorems and Jarzynski's equality,
 see {\it e.g.} \cite{Crooks_2}. Very recently Cleuren, Van den Broeck and Kawai
\cite{cvk} have derived equivalent of Crooks identity in microcanonical ensemble
description and have obtained analytical expressions for the work fluctuations 
in an idealized experiment consisting of a convex body moving 
at constant speed through an ideal gas.  
Crooks identity has since been verified experimentally \cite{CRJSTB}. 

\section{Epilogue}
 Thus,  recent developments have helped improve our understanding 
of the issues that link time asymmetric macroscopic world  to the 
time symmetric microscopic  world.  
These developments  are not inconsistent with 
the hunch  Boltzmann had. 
Let me conclude {\it $\grave{a}$ la} 
Cohen \cite{Cohen_1996}, 
quoting from Boltzmann.
In his 1899 lecture at Munich, Germany, on 
{\it recent developments of methods of theoretical physics} 
\cite{Boltzmann_1899},
 Boltzmann talks of 
the conflict between dynamics and statistics 
in describing macroscopic phenomena. 
He asks  if  
statistics would continue to dominate in the future, or  would it give way to 
dynamics. He concludes saying \lq\ $\cdots$ {\it 
 interesting questions! One almost regrets to have to die long 
before they are settled. Oh! immodest mortal ! 
Your destiny is the joy of watching the ever-shifting battle}\rq\ .
\section*{Acknowledgment}
I must thank amongst many, G. Raghavan, V. S. S. Sastry,
S. L. Narasimhan, V. Sridhar,  L. V. Krishnan,
V. Balakrishnan and J. Srinivasan  
for reading the manuscript and making  helpful comments. 
I have benefited from reading 
several papers and books.  
Some, not listed
under references, are listed  at the end,
 under bibliography.   

\bigskip
\bigskip
{\Large{\bf Bibliography}}
\bigskip
\begin{enumerate}

\item[$\bullet$]
P. Ehrenfest and T. Ehrenfest, {\it The conceptual foundation of the 
statistical approaches in mechanics}, Cornell University Press, 
New York (1959); translated from papers published in 1912.

\item[$\bullet$]
Max Planck, {\it Treatise on Thermodynamics}, Dover Pub. Inc., translated from the 
Seventh German edition (1922). 

\item[$\bullet$]
K. K. Darrow, {\it The concept of Entropy}, 
Am. J. phys. {\bf 12}(4), 183, Aug. (1944).

\item[$\bullet$]
E. Mendoza (Editor), {\it Reflections on the Motive Power of
Fire and Other Papers}, Dover Publications Inc., New York (1960).
 
\item[$\bullet$]
Max Born, {\it Natural Philosophy of Cause  and Chance}, 
(The Waynflete Lectures
delivered at the College of St. Mary Magdalen, oxford in Hilary term,
in 1948), University press, Oxford (1949). 

\item[$\bullet$]
H. Rief, {\it Fundamentals of Statistical and Thermal Physics},
McGraw Hill, New York (1965).

\item[$\bullet$]
E. B. Stuart, B. Gal-Or and A. J. Brainardd (Editors), {\it A Critical Revioew of Thermodynamics},
Mono Book Corp., Baltimore (1970).

\item[$\bullet$]
O. Penrose, {\it Foundations of Statistical mechanics},
Pergamon, New York (1970).

\item[$\bullet$]
Jacques Monod, {\it chance and Necessity}, Wainhouse, Austryn (1971).

\item[$\bullet$]
 E. G. D. Cohen and W. Thirring (Eds.), {\it The Boltzmann equation,
theory and applications}, Springer (1973).

\item[$\bullet$]
R. Balescu, {\it Equilibrium and Non-equilibrium statistical mechanics},
Wiley (1975).

\item[$\bullet$]
I. Prigogine, {\it From Being to becoming}, Freeman, San Francisco (1980).

\item[$\bullet$]
P. Atkins, {\it The Second law}, W. H. Freeman and co., New York (1984).

\item[$\bullet$]
H. B. Callen, {\it Thermodynamics}, John Wiley (1985).

\item[$\bullet$]
D. Chandler, {\it Introduction to modern statistical mechanics},
Oxford University Press, New York (1987).

\item[$\bullet$]
B. Holian, W. Hoover and H. Posch, {\it Resolution of Loschmidt's 
paradox: the origin of irreversible behaviour in 
reversible atomistic dynamics}, Phys. Rev. Lett. {\bf 59}, 10 (1987).

\item[$\bullet$]
Kersen Huang, {\it Statistical Mechanics}, Wiley Eastern, New Delhi (1988). 

\item[$\bullet$]
M. C. Mackay, {\it The dynamical origin of increasing entropy},
Rev. Mod. Phys. {\bf 61}, 981 (1989).

\item[$\bullet$]
William G. Hoover, {\it Computational Statistical Mechanics},
Elsevier, Amsterdam (1991); can be downloaded from
http://williamhoover.info/book.pdf

\item[$\bullet$]
D. Ruelle, {\it Chance and Chaos}, Princeton University press, Princeton (1991).

\item[$\bullet$]
M. Schroeder, {\it Fractals, Chaos and Power laws: 
Minutes from an infinite paradise},
W. H. Freeman and Co., New York (1991).

\item[$\bullet$]
E. G. D. Cohen, {\it Fifty years of kinetic theory}, Physica A {\bf 194}, 229 (1993).

\item[$\bullet$]
J. Lebowitz, {\it Boltzmann's Entropy and Time's Arrow},
Physics Today, {\bf 46} 32,  Sept. 1993.

\item[$\bullet$]
J. L. Lebowitz, {\it Macroscopic laws and Microscopic Dynamics, 
Time's arrow and Boltzmann's entropy}, 
Physica A{\bf 194}, 1 ((1993) 

\item[$\bullet$]
R. Baierlin, {\it Entropy and the Second law: A pedagogical approach},
Am. J. Phys. {\bf 62} (1), 15,  Jan. (1994).

\item[$\bullet$]
G. Gallavotti, {\it Ergodicity, ensembles, irreversibility in Boltzmann and beyond},
arXiv: chao-dyn/9403004 v1 28 March 1994.

\item[$\bullet$]
C. Garrod, {\it Statistical Mechanics and Thermodynamics},
Oxford University Press, New York (1995).

\item[$\bullet$]
M. Guillen,  An unprofitable experience: Rudolf Clausius and the 
Second law of thermodynamics, p. 165 in {\it Five Equations that 
changed the world: the power and poetry of Mathematics},
Hyperion, New York (1995).

\item[$\bullet$]
R. K. Pathria, {\it Statistical Mechanics} Butterworth-Heinemann, 
Oxford
(1996).

\item[$\bullet$]
D. Flamm, {\it History and outlook of Statistical Physics},
UWThPh-1997-52, 27 Oct. 1997: arXiv: physics/9803005 v1 4 March 1998.

\item[$\bullet$]
D. Flamm,
{\it Ludwig Boltzmann -- A Pioneer of Modern Physics},
arXiv: physics/9710007  

\item[$\bullet$]
H. S. Leff, {\it What if entropy were dimensionless ?}, Am. J. Phys. 
{\bf 67}(12), 1114, Dec. (1999).

\item[$\bullet$]
William G. Hoover, {\it  Time Reversibility, Computer Simulation,
and Chaos}, World Scientific, Singapore (1999).

\item[$\bullet$]
J. L. Lebowitz, {\it Statistical Mechanics: a selected review of two
central issues}, Rev. Mod. Phys. {\bf 71}, S346 (1999).

\item[$\bullet$]
I. Prigogine, {\it law of nature, probability and time symmetry breaking},
Physica A{\bf 263}, 528 (1999).

\item[$\bullet$]
D. J. Searles, and D. J. Evans, {\it The fluctuation theorem and Green-Kubo relations},
arXiv: cond-mat/99022021 (1999).

\item[$\bullet$]
D. F. Styer, {\bf Insight into Entropy}, Am. J. Phys. {\bf 68}(12), 
1090, Dec. (2000).

\item[$\bullet$]
S. R. A. Salinas, {\it Introduction to Statistical Physics}, Springer New York (2001).

\item[$\bullet$]
A. M. Glazer and J. S. Wark, {\it Statistical mechanics: A Survival Guide},
Oxford University press, Oxford (2001). 
 
\item[$\bullet$]
David Lindley, {\it Boltzmann's Atom: The Great Debate that 
Launched a Revolution in Physics}, Free Press, New York, 2001. 

\item[$\bullet$]
C. Callender, {\it Thermodynamic Asymmetry in Time},
The Stanford Encyclopedia of Philosophy (Winter 2001 Edition), 
Edward N. Zalta (ed.),\\ 
URL:  http://plato.stanford.edu/archives/win2001/entries/time-thermo/.  

\item[$\bullet$]
S. Goldstein, {\it Boltzmann's approach to statistical mechanics}, 
arXiv: cond-mat/0105242 v1 May 2001.

\item[$\bullet$]
J. Srinivasan, {\it Sadi Carnot and the Second law of thermodynamics},
Resonance {\bf 6}(11), 
42 (2001).

\item[$\bullet$]
B. J. Cherayil, {\it Entropy and the direction of natural change},
Resonance {\bf 6}(9), 82 (2001)

\item[$\bullet$]
J. K. Bhattacharya, {\it Entropy $\grave{{\rm a}}$ la  Boltzmann},
Resonance {\bf 6}(9), 19 (2001).

\item[$\bullet$]
D. C. Schoepf, {\it A statistical development of
entropy for introductory physics course}, Am J. Phys. {\bf 70} (2),
128, Feb. (2002). 

\item[$\bullet$]
F. Ritort, {\it Work fluctuations and transient violations of the 
Second law: perspectives in theory and experiments}, 
S\'eminaire Poincar\'e
{\bf 2}, 63 (2003). 

\item[$\bullet$]
Michel le Bellac, Fabrice Mortessagne and G. George 
Batrouni, {\it Equlibrium and non-equilibrium statistical thermodynamics},
Cambridge University Press (2004).

\item[$\bullet$]
Y. Oono, {\it Introduction to non-equilibrium statistical thermodynamics}, 
version - 0.5 (2005).

\end{enumerate}
\end{document}